\documentclass[reprint,aps,prl]{revtex4-1}

\newcommand*{\Sequence}{\ensuremath{\mathcal{R}}}
\newcommand*{\Structure}{\ensuremath{\mathcal{S}}}

\usepackage{graphicx}
\usepackage{amsmath}
\usepackage{amsfonts}
\usepackage{amssymb}

\graphicspath{{data/}}

\begin{document}
\title{Extremely rare ultra-fast non-equilibrium processes
  can be close to equilibrium:
  RNA unfolding and refolding}
\author{Peter Werner}
\author{Alexander K. Hartmann}	
\affiliation{Institut f\"ur Physik, Universit\"at Oldenburg, 26111 Oldenburg,
  Germany}
\begin{abstract}
We study numerically the behavior of RNA secondary structures under
influence of a varying external force. This allows to measure the work
$W$ during the resulting fast unfolding and refolding processes.
Here, we investigate a medium-size hairpin structure.
Using a sophisticated large-deviation algorithm,  we are able to
measure work distributions with high precision 
 down to probabilities as small as $10^{-46}$. Due to this precision and by
comparison with exact free-energy calculations we are able to verify
the theorems of Crooks and Jarzynski.
Furthermore, we analyze force-extension curves and the
configurations of the secondary structures during unfolding and
refolding for typical equilibrium processes and non-equilibrium processes,
conditioned to selected values of the measured work $W$, typical
and rare ones.  We find that the 
non-equilibrium processes where the work values are close to those
which are most relevant for applying Crooks and Jarzynski theorems,
respectively, are most and quite similar to the equilibrium
processes. Thus, a similarity of equilibrium and non-equilibrium
behavior with respect to a mere scalar variable, which occurs with a
very small probability but can be generated in a controlled but non-targeted
way, is related to a high similarity for the set of configurations sampled
along the full dynamical trajectory.
\end{abstract}

\maketitle

In Statistical Physics, the cleanest and so far best-justified description
is obtained for systems in equilibrium.
Nevertheless, due
to open system boundaries and lack of infinite time to perform
experiments or simulations, most real and simulated
model systems are constantly in
non-equilibrium. A bridge between both worlds is provided.
e.g., by the theorems of Jarzynski \cite{jarzynski1997} and Crooks
\cite{crooks1998}, where the distribution $P(W)$
of work $W$ is measured for arbitrary
fast non-equilibrium processes
obtained from sampling
equilibrium initial configurations and possibly stochastic
non-equilibrium trajectories. Correspondingly $P_{\rm rev}(W)$ is the
distribution for the reverse process. For a system coupled
to a heat bath, Crooks theorem reads
$P(W)=P_{\rm rev}(-W)\exp(-(\Delta F-W)/T)$. This can
be used to reconstruct the true free
energy difference $\Delta F$ between initial and final state,
because $P(W)$ and $P_{\rm rev}(-W)$ cross at $W=\Delta F$.
Correspondingly the equation of Jarzynski reads 
$\langle e^{-W/T}\rangle= e^{-\Delta F/T}$.
These and related theorems have lead
to many applications and extensions  relating
equilibrium and non-equilibrium processes
\cite{kurchan2007,seifert2008,sevick2008,jarzynski2008,esposito2009,jarzynski2011,seifert2012,marsland2018}.
A fruitful field of applications is biophysics, where these theorems
are used to measure properties of small molecules like RNA.

One major goal of stochastic thermodynamics is to extract
equilibrium information from non-equilibrium measurements or
simulations \cite{practical_guide2015}.
The  fluctuation theorems concern specific measurable
scalar quantities like work \cite{jarzynski1997,crooks1998,hummer2010},
entropy \cite{evans1993,evans1994,gallavotti1994,gallavotti1995,kurchan1998,lebowitz1999,maes1999,crooks1999,crooks2000},
or a quantity measuring the volume of the phase space  
\cite{adib2005}.
However, beyond statistics of particular scalar quantities, the fluctuation
theorems do not provide information about
the unseen equilibrium behavior along the trajectory, i.e.,
with respect to arbitrary measurable quantities.
Standard derivations of the fluctuations theorems
only involve terms which include energies and probabilities of the
initial and final state.
What may we expect when we analyze the full trajectory
of a non-equilibrium process?
First, a \emph{typical}, i.e., highly probable
sample of a non-equilibrium trajectory
will look very different from a corresponding trajectory
sampled during an equilibrium process.
Second, it is known that
when reweighting trajectories suitably in a time-dependent
way, they also carry some
information about the intermediate not-seen equilibrium
states \cite{jarzynski1997long,crooks2000,hummer2001} which
allows for the reconstruction of full free-energy profiles beyond
initial and final state.
Third, it is somehow intuitive to believe that the \emph{rare}
non-equilibrium processes
which contribute most to the estimation of $\Delta F$ are in a
comprehensive way, without reweighting,
similar or even equal to the corresponding equilibrium processes.
For the case of the theorems of Crooks and Jarzynski,
the statistics of the work distributions are most relevant for
particular work values $W=\Delta F$ and $W=W_{\rm J}^*$, where the
latter one is the
value where the integrand $e^{-W/T}P(W)$ exhibits a maximum. Note that
these values are high improbable to occur for large system sizes.
On the other hand, beyond this intuition, there is no solid
reason that these rare possibly very fast processes completely resemble
true equilibrium processes: 
A non-equilibrium process
always depends on the history, i.e., on
many configurations encountered so far,
while each equilibrium state in a process
does not depend at all on the history. In particular,
non-equilibrium processes depend
on the speed of performance, while the equilibrium is for infinite low
speed. 

This question motivates our present work:
We investigate in a comprehensive way the
dynamics of fast non-equilibrium processes conditioned to various
non-equilibrium work values $W$, typical and rare ones,
and compare with the equilibrium process behavior.
In particular, we study unfolding and refolding of RNA secondary
structures subject to an external force \cite{mueller2002}.
The former one, denoted as \emph{forward} process,
involves stretching an RNA by subjecting it to
an external force $f$ which is increased from starting at zero.
For the latter
one, denoted as \emph{reverse} process,
one starts with a large force and reduces it to zero.
For small RNAs consisting of few dozens of bases, 
Crooks theorem has been confirmed in experiments
and simulations \cite{hummer2010} for slow unfolding and
refolding processes. For such small RNA
and slow processes,
the resulting work distributions are rather broad and the  
distribution for forward and reverse processes are close to each
other such that they cross at high-probability
values which are easily accessible.
For larger RNA molecules, the crossing points will move to smaller
probabilities, such that the crossing cannot be observed in experiments
or standard simulations. 
To go beyond such limiting system sizes, we applied for our study
sophisticated large-deviation algorithms \cite{align2002,bucklew2004},
which allowed us
to measure probability distributions numerically down to extremely small
probabilities. These algorithms have also applied successfully
to non-equilibrium processes like the transition-path sampling
approach to study protein folding \cite{dellago1998,bolhuis2002},
population-based
approaches to study
asymmetric exclusion processes \cite{giardina2006,lecomte2007}
or Markov-chain Monte Carlo methods to
investigate, e.g., traffic models \cite{nagel_schreckenberg2019}
and the Kardar-Parisi-Zhang equation \cite{kpz2018}.
In particular such an algorithm has also been applied to measure
with high precision
the work distribution of an Ising model subject to a varying external field
\cite{work_ising2014},
providing the first confirmation of the theorems of Jarzynski and Crooks
for a large system with many thousands of particles.

Thus, here we will provide similar evidence for RNA secondary
structure unfolding and refolding
by applying such a rare-event algorithm, allowing us to obtain the
work distributions of intermediate-sized RNAs down to probabilities
as small as $10^{-46}$. Furthermore, we will
analyze the temporal structure of the non-equilibrium processes,
conditioned to the occurring work values $W$. We will compare
this to the corresponding equilibrium process,
which can be sampled exactly \cite{higgs1996,stacks2005,lorenz2011}
and efficiently, i.e., in polynomial time,
for RNA secondary structures without pseudo-knots. Beyond confirming
the theorems of Jarzynski and Crooks we find in particular that the
non-equilibrium processes can be very similar in their
development
to the equilibrium ones. The highest similarity is reached
for processes which exhibit a work value $W$ \emph{between} the 
values $W=\Delta F$ and $W=W_{\rm J}^*$ which are most relevant for
the Crooks and Jarzynski theorem, respectively.

We will next present our model and the simulation methods we used. Then
we show our results and finish by a discussion.

\emph{Model} ---
Each RNA molecule is a linear chain of length $L$ of bases from
$\{\mathrm{A,C,G,U}\}$. A \emph{secondary structure} is a set of pairs
of bases, such that only complementary (Watson-Crick) base pairs
A-U and C-G are allowed. We forbid \emph{pseudo-knots}, which means
that it is always possible to draw the molecule as a single line
and connect all pairs by lines such that no intersections occur.

The energy of an RNA secondary structure consists, first, of the energy
from the Watson-Crick pairs, which is its number here for simplicity.
Second, the RNA is subject to a force $f$. This gives rise to an energy
contribution $-f\times n$ where $n$ is the \emph{extension}
of the structure, i.e., the part of the RNA which is outside
any paired base, plus length 2 for any paired base on the
first level.
For details see the supplementary material
(SM).

\emph{Algorithms} ---
For RNA secondary structures it is possible to
sample them directly in equilibrium for finite
temperatures $T$ in time $O(L^3)$.  We used an extension
of the approach for the zero-force case
\cite{higgs1996}. For this purpose, one needs also to calculate
partitions functions for some sub sequences, which is possible
using \emph{dynamic programming} in polynomial time. 
These approaches  \cite{gerland2001,mueller2002}
are also  extensions of the 
zero-force case approach \cite{nussinov1978}. For details see the SM.

For actually performing an unfolding or refolding process, and to
measure the performed work $W$, we started
with a configuration $\mathcal{S}_0$ 
sampled in equilibrium at initial force $f=f_0$ with $f_0=0$ or
$f_0=f_{\max}=2$. 
 Then the force $f$  was gradually changed in steps $\Delta f$ up to $f_{\max}$
or down to zero, respectively, while
allowing for thermal fluctuations by performing Monte Carlo simulations
with creation or removal of pairs as basic  moves.  Each time the force 
is changed, we
obtained a contribution $\Delta W = -n(\mathcal{S})\Delta f$ to the work, where
$n(\mathcal{S})$ is the current extension.
For details of the algorithm see the SM.

By repeating an unfolding or refolding simulation many times,
one can measure approximately the work distributions $P(W)$ and
$P_{\rm rev}(W)$, respectively. Nevertheless, this \emph{simple
  sampling approach} allows one only to obtain the work distributions
down to rather large probabilities, like $10^{-9}$. To obtain the
work distributions down to much smaller probabilities, we applied
sophisticated large-deviation algorithms \cite{align2002,bucklew2004}.
Our approach has already
been used to measure work distributions for large Ising systems
\cite{work_ising2014}. The basic idea is to drive the
forward and reverse processes, respectively,
by vectors $\xi$ of random numbers and control the composition
of the vectors with a Markov chain Monte Carlo simulation, with
a known, i.e., removable, bias depending on the measured work.
For details see the SM.

\emph{Results} ---
We considered an RNA sequence which is not too small, such that we
were  able to
observe differences between equilibrium and non-equilibrium secondary
structure configurations
with suitable resolution. We studied a hairpin structure of length
$L=100$ which has a sequence (AC)$^{25}$(UG)$^{25}$, resulting in a ground
state of one large stack with a small hairpin.
For such an RNA size  the application of
large-deviation algorithm is necessary to measure the work distribution
with suitable accuracy such that the theorems of Jarzynski and Crooks
can be applied and the unfolding and refolding histories captured.
We considered the RNA coupled to a heat bath
at temperatures $T=0.3$ and $T=1$, respectively. These are
low enough temperatures, such that in the force-free case,
the RNA is basically folded, but exhibits thermal fluctuations.
Example secondary structures are shown in a figure in the SM.
We considered two different speeds of the folding processes, i.e.,
two different numbers $n_{\rm MC}$ of sweeps performed during
the process, here $n_{\rm MC}=8$ and $n_{\rm MC}=16$.
A summary of
the simulation parameters is also given in the SM.

\begin{figure}
  \includegraphics[width=\linewidth]{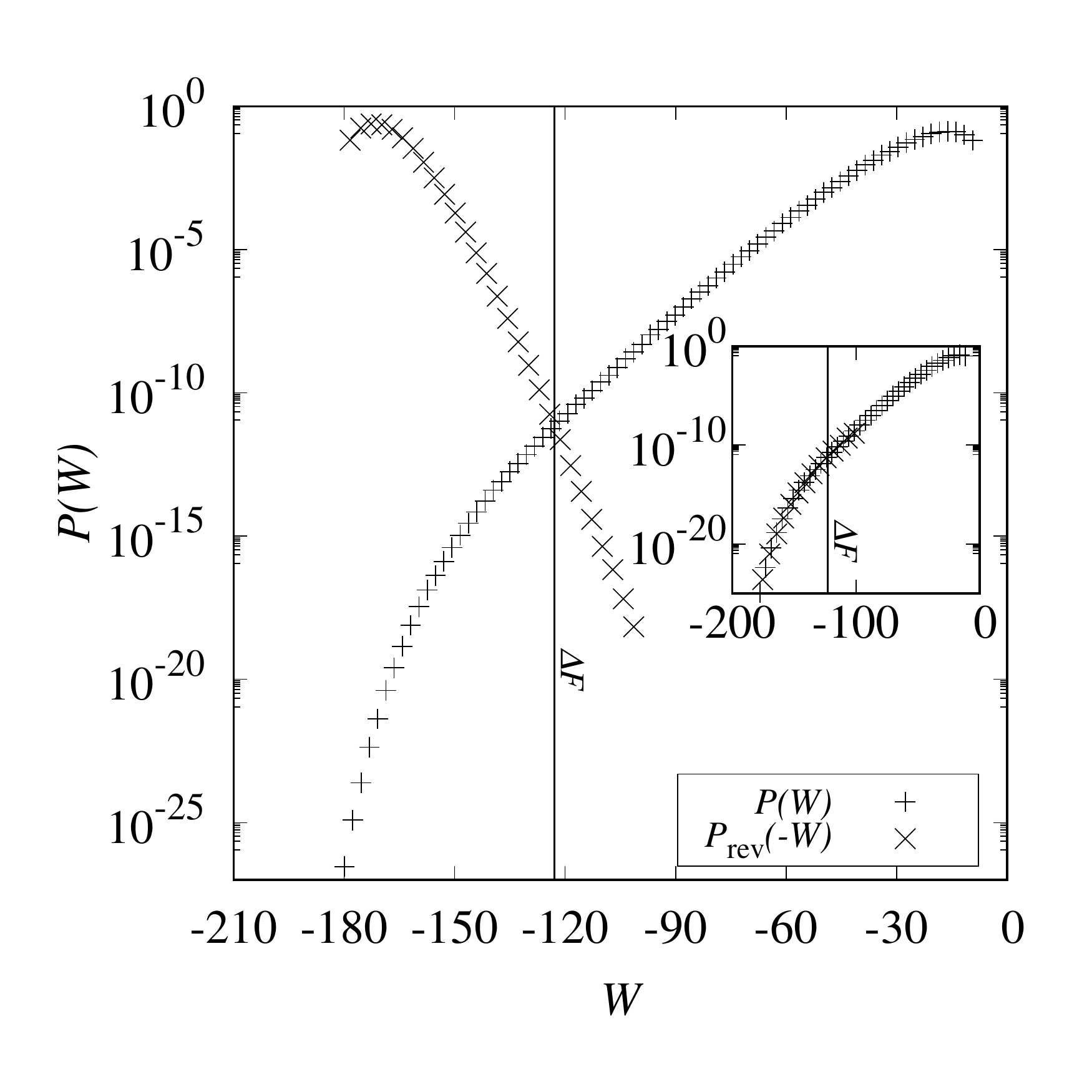}
	          \caption{\label{fig:workdistribution} Plain and mirrored
                    work distributions for
          $T=1$ and 8 sweeps of the forward and reverse process, respectively.
                    They intersect near $W=\Delta F$, which
          is the exact value and indicated by the vertical line.
          The inset shows the same plot but with the distribution for the
          reverse process (cross symbols) 
          rescaled as $P_{\rm rev}(W)\exp(-(\Delta F-W)/T)$,
          according to the equation of Crooks, yielding a good
                  agreement with $P(W)$.}
\end{figure}

In Fig.~\ref{fig:workdistribution} the work distributions $P(W)$
of the forward and $P_{\rm rev}(-W)$ of the reverse 
processes are shown for the case $T=1$ and $n_{\rm MC}=8$.
With the application of the large-deviation scheme,
we are able to resolve very small probabilities down
to $10^{-26}$, i.e., over 26 orders in magnitude.
The crossing of the distributions at a work value $W=\Delta F$
predicted by the theorem of Crooks \cite{crooks1998} can be well observed.
For the present model, because we
can exactly calculate numerically the partition function, we are
able to obtain
$\Delta F = 1/T \log \{Z(f=f_0)/Z(f=f_{\max})\}$.  Apparently the data
matches the expectations from Crooks theorem with high precision.

Crooks relation means that when $P_{\rm rev}(-W)$ is rescaled according
the exponential, it equals $P(W)$. This is also confirmed very convincingly
by our data
over up to 20 decades, as shown in the inset of 
Fig.~\ref{fig:workdistribution}. This in particular shows
\cite{work_ising2014}
that our higher-level MCMC simulation is well equilibrated. We obtained
similar results
for the slower $n_{\rm MC}=16$ process. We also studied
the lower temperature $T=0.3$ with $P(W)$ even down to $10^{-46}$, see the SM.

Our results allow us to go
beyond calculation of distributions and study the actual dynamic processes,
conditioned to any value of $W$. We concentrate now on $T=1$,
the results for $T=0.3$ are similar. During a forced process, we sampled
structures, one for each considered value of $f$,
in equilibrium and in
non-equilibrium. To compare two sampled
structures we define an \emph{overlap} $\sigma$,
which runs over all bases of the sequence, and counts $1/L$ if for
both structures the base is not paired or if for both structures
it is paired with the same base. Otherwise zero is counted, see SM for
a formal definition. Overlaps quantities are used frequently to
determine order in complex systems, e.g., spin glass \cite{mezard1987}.

\begin{figure}
  \includegraphics[width=0.49\linewidth]{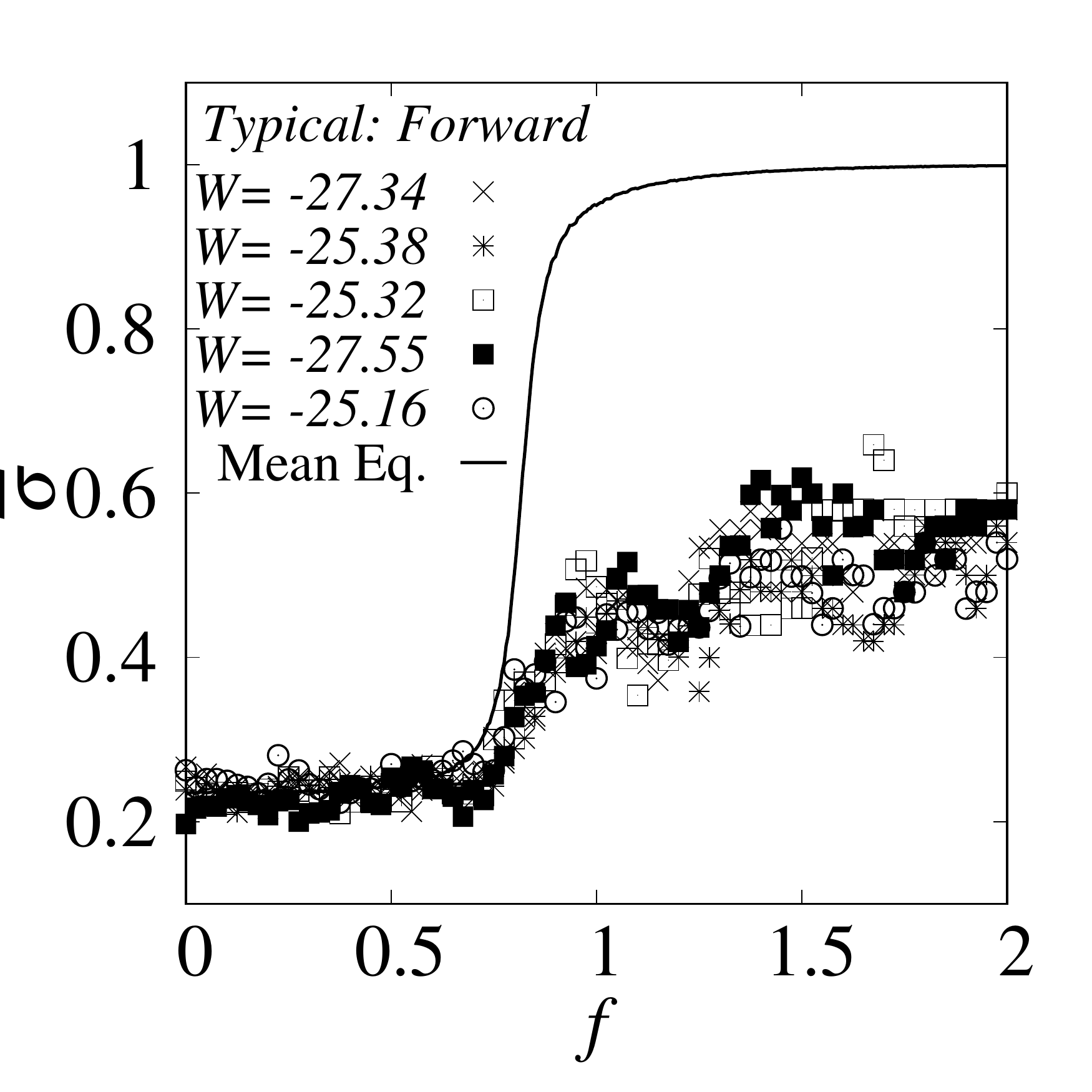}
  \includegraphics[width=0.49\linewidth]{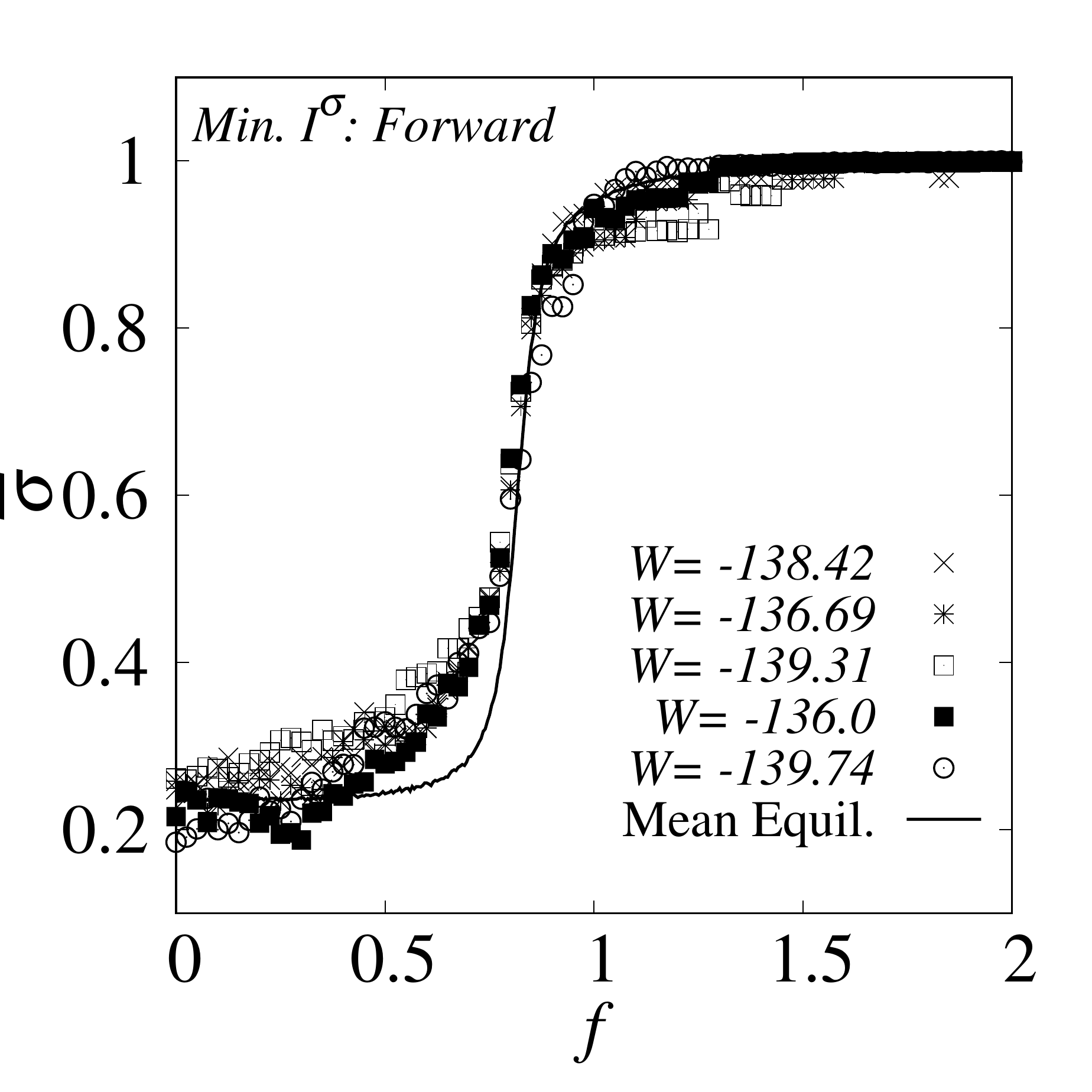}
  \includegraphics[width=0.49\linewidth]{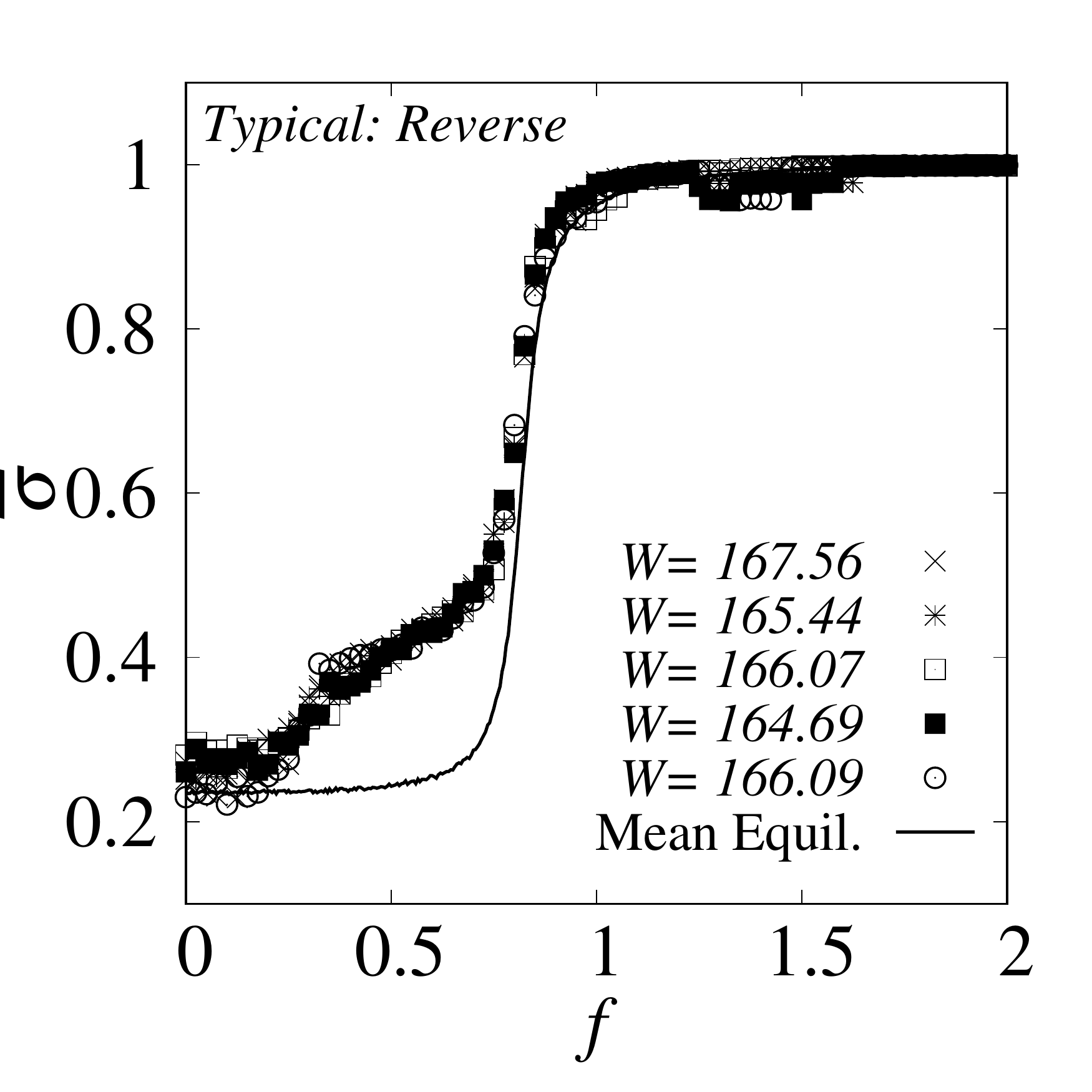}
  \includegraphics[width=0.49\linewidth]{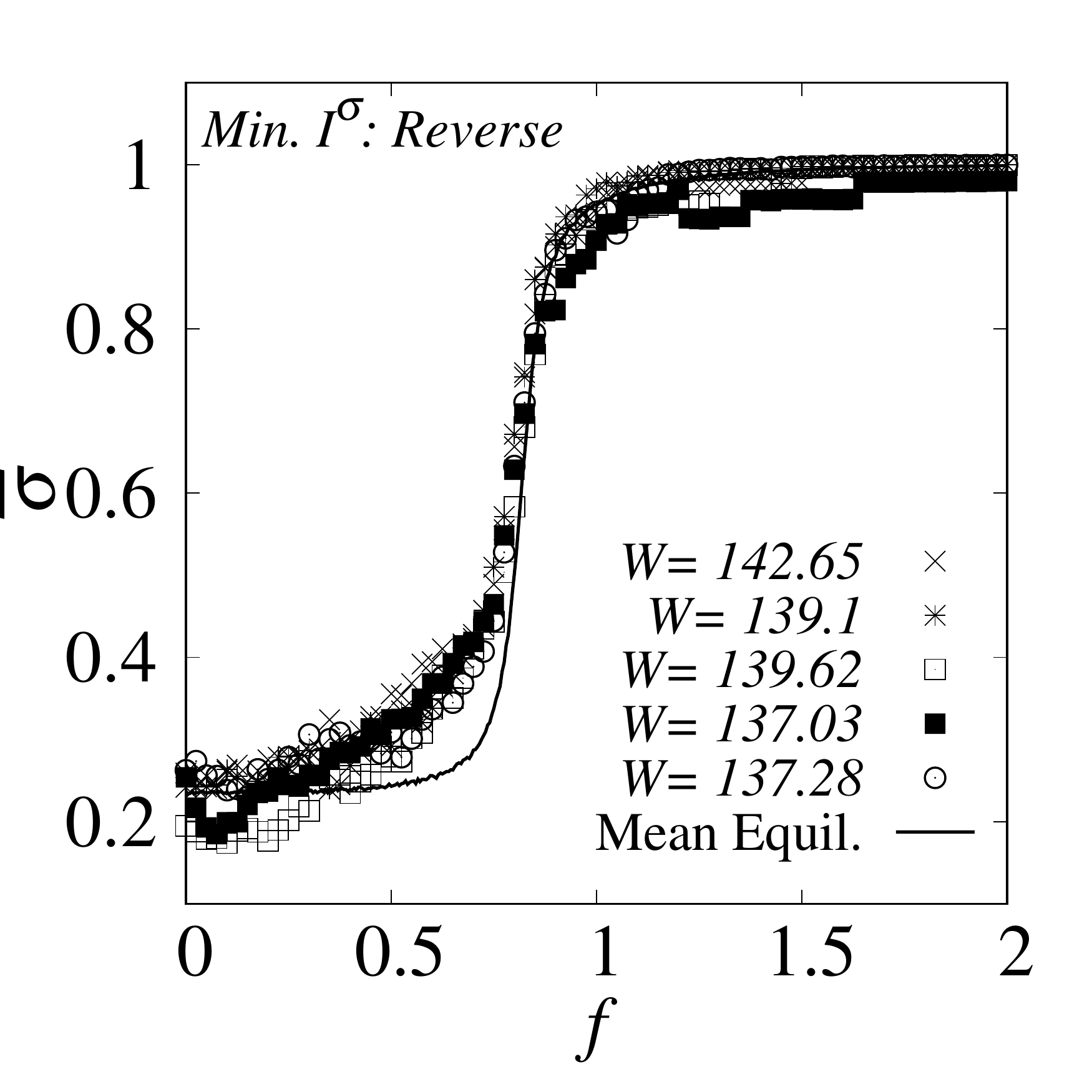}
  \caption{\label{fig:overlap:curves} 
    Average non-equilibrium overlap profiles $\overline{\sigma}(f)$, i.e. between one
    non-equilibrium and one equilibrium structure,
    for some sample
    processes at $T=1$ and 16 sweeps, with mentioned non-equilibrium work
    values $W$. The solid line is the averaged
    equilibrium overlap, i.e., between
    two equilibrium structures, respectively.
    Top row: forward process
    for typical (left) and very rare (right) values of $W$.
    Bottom row: the same for the reverse process. Error bars are smaller
    than symbol size. }
\end{figure}

Fig.~\ref{fig:overlap:curves} shows 
average  \emph{non-equilibrium profiles}
$\overline{\sigma}(f)$, i.e., averaged overlaps $\sigma$ as function of $f$,
where in the calculation of the overlaps
one structure is a given non-equilibrium
sample of a forward or a reverse
process  and the other structure
is a sampled equilibrium structure. Always an
average is taken over many equilibrium structures.
For comparison in all plots
the average \emph{equilibrium}
profile is shown, where both structures are sampled
from equilibrium.  Our results show that folded structures at low force value
$f$ are characterized by a variety of secondary structures, while
at high values of $f$, where the RNA is basically stretched,
the secondary structures are very similar to each other.
We see that for \emph{typical}
work values, i.e., where $P(W)$ and $P_{\rm rev}(W)$ peak,
in particular for the forward process,
large differences for non-equilibrium 
profiles compared to the average equilibrium profile occur. For 
work values near $W=\Delta F\approx -123$ on the other hand, we observe a high
similarity, i.e., these very rare non-equilibrium
processes enroll close to the equilibrium ones.

\begin{figure}
	\includegraphics[width=\linewidth]{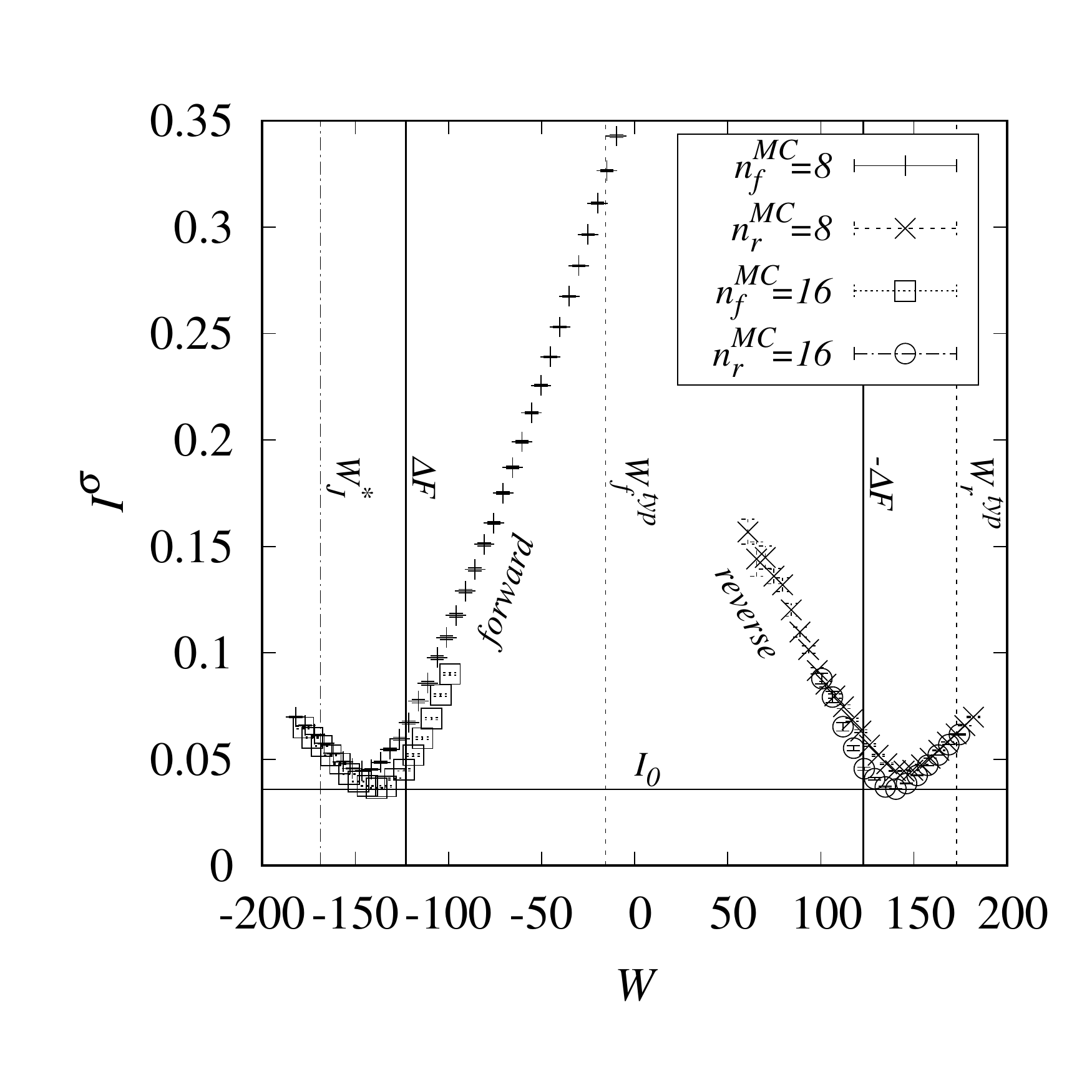}
\caption{\label{fig:integrated:overlap:difference} Integrated
  difference $I^{\sigma}$ between equilibrium and
  non-equilibrium overlap profiles at $T=1$, for forward (left)
  and backward (right) processes. For 16 sweeps the data is only
  partially shown, for better visibility.
  The horizontal line indicates $I_0$, 
  the value of $I^{\sigma}$ in equilibrium.
  Vertical lines indicate work values at (from left
  to right) the maximum $W_{\rm
    J}^*$ of the Jarzynski integrand, the free energy difference $\Delta F$,
  the maximum $W^{typ}_f$ of the forward process work distribution, the
  negative free energy difference $-\Delta F$ and the maximum
  point $W^{typ}_r$ of the reverse process work distribution.
}
\end{figure}

We quantify the similarity $I^{\sigma}$
of the non-equilibrium processes to the equilibrium
case by integrating over all force values $f$ the absolute difference
of $\overline{\sigma}(f)$ between the equilibrium and non-equilibrium case, and
average this integral over close-by values of $W$, i.e.,
obtaining $I^{\sigma}(W)$. The result is shown in
Fig.~\ref{fig:integrated:overlap:difference}. We observe
rather larger differences for typical values of $W$, while
near $W\approx \pm \Delta F$ the similarity is of the order
of the similarity $I_0$ obtained by averaging $I^{\sigma}$
over many independent
equilibrium processes, which represents the equilibrium fluctuations.
Also the forward processes sampled for work
values near the value $W_J^*\approx -170$ where the 
Jarzynski integrand $P(W)e^{-W/T}$ peaks (see SM) exhibit a high similarity
to the equilibrium case. Note that for the reverse process, the
value of $W_J^*$ occurs outside our sampled region, thus we do not have
processes for this case. For the slower case of $n_{\rm MC}=16$
sweeps, i.e.,
a bit nearer to equilibrium, the location minimum moves closer to
$\Delta F$ and even decreases in height towards the equilibrium value
$I_0$.

Thus, our results show that the rare processes
near $W=\pm \Delta F$ do not only have
similar work values like the equilibrium processes, they exhibit
also very similar sequences, as function of the force $f$,
of sampled structures. We obtained a similar result 
when considering force-extension curves, see SM.

\emph{ Discussion} ---
We studied RNA unfolding and refolding in equilibrium and in non-equilibrium.
For the non-equilibrium case, by using sophisticated large-deviation
algorithms, we could access
a large range of the support of the probability distribution for the
work. This allowed us to confirm the theorems of Crooks and Jarzynski over
several dozens decades in probability. Furthermore, we analyzed
the trajectories in force-extension as well as in secondary-structure
space conditioned to various values of $W$. We observe that near the
most relevant, but very improbable, values of $W$,
the sampled trajectories reach a high similarity with true equilibrium.
Thus, the study here does not depend
on assigning a time-dependent weight to the trajectories, the selection is
solely by the total work performed during the process and suitably
evaluating fluctuation theorems.
Also no other particular similarity to equilibrium is enforced explicitly
by our procedure.
Our approach and results may open a pathway to learning not only
about equilibrium characteristic scalar numbers from non-equilibrium
measurements, but even investigating
near equilibrium dynamics by performing
very fast but biased non-equilibrium simulations.
We anticipate that similar studies are feasible and useful
for many different types of systems.

For further studies,
one could also extend the approach, by storing the configurations
of the close-to-equilibrium $W\approx \Delta F$
generated rare trajectories. Starting with these
configurations one could perform
additional equilibrium simulations at fixed force values, i.e.,
without performing work, in the hope
to get quickly close or even up to equilibrium.
We have run some test simulations which show that one can indeed
get even much closer to the equilibrium behavior by applying these add-on
equilibration, apparently perfectly with respect to the force-extension
curves, but this also depends on the temperature. Here more studies
are needed,
in particular a comparison of how good one can equilibrate by just using
secondary-structure MC simulations when starting with empty configurations.
Also it would be very interesting to see how these results depend of the
actual RNA sequence and the corresponding  energy landscape.

\acknowledgments{
  The simulations were performed at the 
    the HPC cluster CARL, located at the University of Oldenburg (Germany) and
    funded by the DFG through its Major Research Instrumentation Program
    (INST 184/157-1 FUGG) and the Ministry of Science and Culture (MWK) of the
    Lower Saxony State. 
}

\section{Supplemantary Material}

\subsection{RNA secondary structure model}

Each RNA molecule is a linear chain $\Sequence=(r_i)_{i=1,\dots,L}$
of bases, also called residues, with
$r_i\in\{\mathrm{A,C,G,U}\}$ and  $L$ is the length of the sequence. 
For a given sequence $\Sequence$ of bases the secondary structure can be
described by a set $\Structure$ of pairs $(i,j)$ (with the convention $1\leq
i<j\leq L$), meaning that bases $r_i$ and $r_j$ are paired.
For convenience, we also use $s(i)=j$ if $i$ is paired to $j$,
which implies $s(j)=i$, and
$s(i)=0$ if $i$ is not paired.
We only allow Watson-Crick base pairs. These are formed
by hydrogen bonds between \emph{complementary} pairs of bases, i.e.,
A-U and C-G. Formally, this
means for A-U either $r_i=$A and $r_j=$U or vice versa,
correspondingly for the C-G pair.
Two restrictions are used: 
(i)
We exclude so called \emph{pseudo-knots}, that means, for any
$(i,j),(i',j')\in\Structure$, either $i<j<i'<j'$ or $i<i'<j'<j$ must hold, 
i.e., we follow the notion of pseudo knots being more an element of the
tertiary structure \cite{tinoco1999}.
(ii) Between two paired bases a minimum distance is required: $|j-i|> s$ is
required, granting some flexibility of the molecule (here $s=2$).

Every secondary structure \Structure\ is assigned a certain energy
$E(\Structure)$, we do not explicitly
indicate the dependence on the sequence \Sequence.
This energy is defined by assigning each pair $(i,j)$ a
certain energy $e(r_i,r_j)$ depending only on the kind of bases.

\begin{figure}[!h]
\centerline{\includegraphics[width=0.7\linewidth]{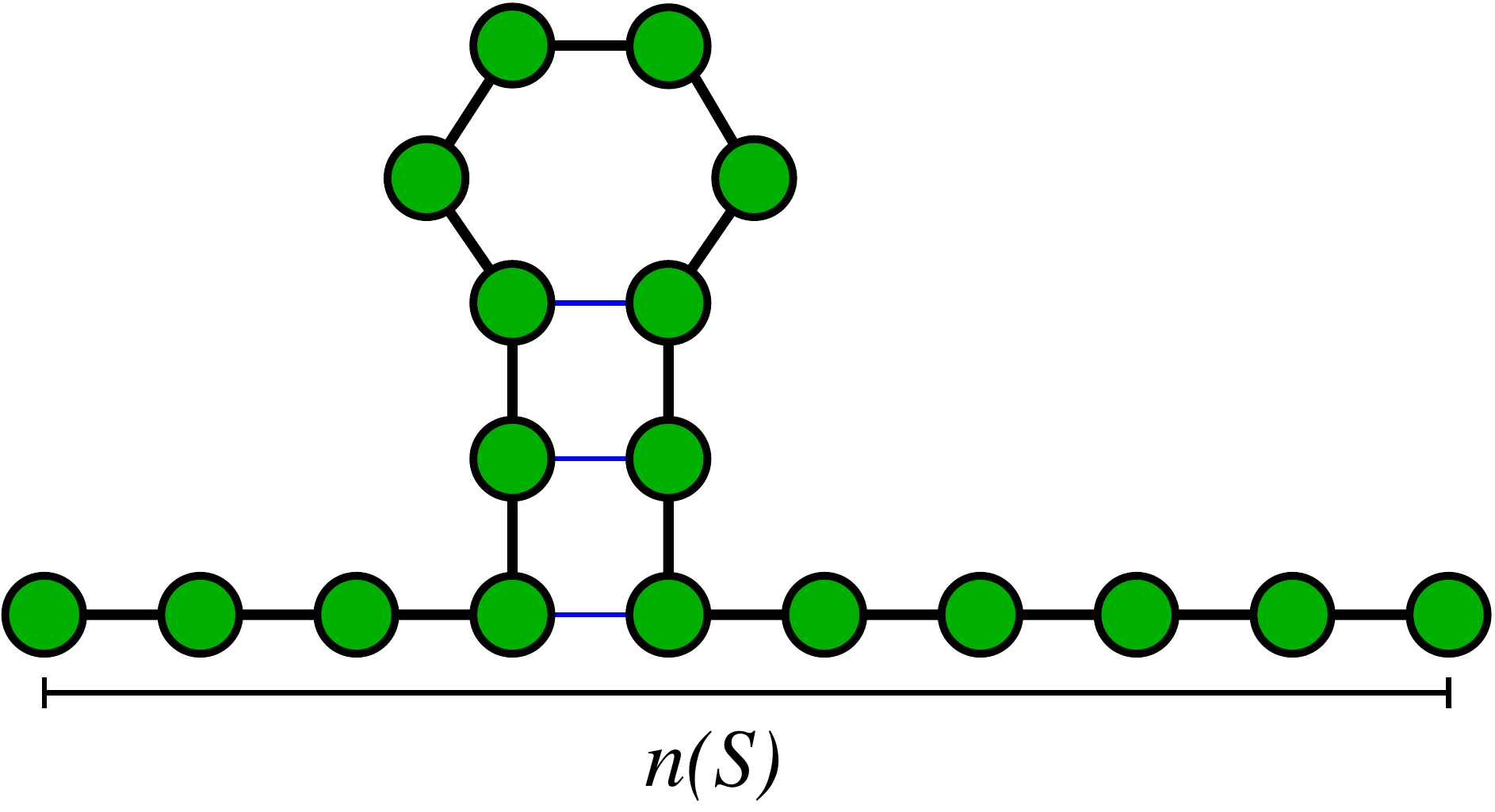}}
\caption{\label{fig:n:S}   (color online)
  An example for a RNA secondary
          structure with one globule and a line indicating the extension
          $n(\mathcal{S})$ of the folded RNA. Circles denote bases,
          thick black lines links between consecutive bases, and thin
        blue lines hydrogen bonds between complementary bases.}
\end{figure}

Furthermore there is a contribution arising from the external force $f$
which stretches the chain to its extension 
$n(\mathcal{S})$, as introduced previously \cite{mueller2002}.
Thus, $n(\mathcal{S})$ is composed of a length of two units for each globule
in the chain plus the number of bases in the free part, i.e., outside
any globule. This is illustrated in Fig. \ref{fig:n:S}.

The total energy is the sum over all pairs plus the interaction with the
external force
\begin{equation}
  \label{eq:paired_energy_model}
E(\Structure)=\sum_{(i,j)\in\Structure}e(r_i,r_j) -n(\mathcal{S})\,f\,.
\end{equation}
 By choosing $e(r, r')=+\infty$ for non-complementary bases
 $r, r'$ pairings of this kind are suppressed. Here we use the most
 simple energy model, i.e., 
 $e(r, r')=-1$ for complementary bases A-U and C-G.

\subsection{Calculation of partition functions}

The partition function $Z_{i,j}$ ($i\le j$) for sub sequence $r_i \ldots r_j$
at inverse temperature $\beta=1/T$ 
without external force and without length constraints,
obeying the minimum distance $s$ between two paired bases, is given by 
\begin{align}\label{eq:partition_function}
		Z_{i,j}&=1 \qquad \text{for} \quad j-i \le  s \nonumber \\
  Z_{i,j}&=Z_{i,j-1}\nonumber\\
  &+\sum_{k=i}^{j-s-1} Z_{i,k-1}
  e^{-\beta e(r_k,r_j)}Z_{k+1,j-1} \qquad  \text{else}
		\end{align}
All $O(L^2)$
values of $Z_{i,j}$ can be conveniently calculated \cite{nussinov1978}
by a \emph{dynamic
programming approach}, i.e. starting
with $Z_{i,i}$ and continuing with increasing values of $j-i$. Since most
contributions involve a sum of $O(L)$ terms, the algorithm
has a running time of $O(L^3)$.

In order to include the interaction with the external force, one needs
additionally the partition function $Q_{1,j,n}$ of the sub sequence 
$r_1 \ldots r_j$
 such that the extension is fixed to the value $n$, with $n \le j$.
We include the fixed index 1 
for matching with the notation for $Z_{i,j}$,

Our approach follows the lines of a corresponding methods
\cite{gerland2001,mueller2002}
for calculation of partition functions and ground state energies of
RNA secondary structures subject to
an external force. 
The partition function reads :\\

\begin{align}\label{eq:free_part_partition_function}
  Q_{1,1,1}=&1 \nonumber \\
  Q_{1,j,1}=&0 \qquad \text{for} \quad j>1, \nonumber\\
  Q_{1,2,2}=&Z_{1,2} \nonumber\\
  Q_{1,j,2}=&0  \qquad \text{for} \quad 2 < j \le s+1\\
  Q_{1,j,2}=& e^{-\beta e(r_1,r_j)}Z_{2,j-1}
  \qquad \text{for} \quad j > s+1 
  \nonumber\\
  Q_{1,j,n}=&Q_{1,j-1,n-1}+\nonumber \qquad\qquad \text{for}\quad  n>2,j\ge n\\
  & \sum_{k=n-1}^{j-s-1} Q_{1,k-1,n-2}
  e^{-\beta e(r_k,r_j)}Z_{k+1,j-1} \nonumber \,.
\end{align}
		
Also all these partition functions can be conveniently calculated by dynamic
programming in time $O(L^3)$.

This allows us to calculate the partition function with force for sub sequence
$r_1,\ldots,r_j$ by
                
\begin{equation}\label{eq:partition_function_with_force}
  \tilde{Z}_{1,j}(f)=\sum_{n=1}^{j}Q_{1,j,n}e^{\beta n f}\,.
\end{equation}
Note that the case $n=0$ can not occur and the case $n=1$
corresponds only to one single base.

\subsection{Sampling secondary structures}

The availability of the above partition functions allows us to
sample secondary structures in the presence of an external force
directly, i.e. rejection free, also in polynomial time.
The approach is an extension of the zero-force algorithm
\cite{higgs1996} to the case $f \ge 0$.

For sampling a structure, the following probabilities are needed.
The probability $p_{i,j,k}^{p}$ that for sub sequence $r_i,\ldots,r_j$,
without the presence or influence of a force, base $j$ is paired
to base $k$ with $i\le k< j$ and $j-k >s$ is given by

\begin{equation}\label{eq:pair_prob_g}
  p_{i,j,k}^{p}=\frac{Z_{i,k-1} e^{-\beta e(r_k,r_j)}Z_{k+1,j-1}}{Z_{i,j}}\,.
\end{equation}
For $j-k\le s$, this probability is zero.
The probability that  base $j$  is not paired is given by
\begin{equation}\label{eq:pair_prob_u}
  p_{i,j}^{u}=\frac{Z_{i,j-1}}{Z_{i,j}}\,.
\end{equation}

The probability $\tilde{p}_{1,j,k}^p(f)$ that for sub sequence $r_1,\ldots,r_j$,
with the presence of a force $f$, base $j$ is paired
to base $k$ with $1\le k< j$ and $j-k >s$ is given by
\begin{equation}\label{eq:pair_prob_force_g}
  \tilde{p}_{1,j,k}^p(f)=\frac{\tilde{Z}_{1,k-1}(f) e^{-\beta e(r_k,r_j)+\beta 2 f }Z_{k+1,j-1}}{\tilde{Z}_{1,j}(f)}
\end{equation}
For $j-k\le s$, this probability is zero.
The probability that  base $j$  is not paired is given by
\begin{equation}\label{eq:pair_prob_force_u}
  \tilde{p}_{1,j}^u(f)=\frac{\tilde{Z}_{1,j-1}(f)e^{\beta f}}{\tilde{Z}_{1,j}(f)}.
\end{equation}

The sampling of a structure is now performed as follows. 
Each time one starts for the full 
sequence $r_1,\ldots,r_L$ by considering the \emph{case with force} $f$:
\begin{itemize}
\item \emph{Case with force $f$} for sub sequence $r_1,\ldots,r_j$

Base $j$ is paired to one of the bases $k=1,\ldots,j-s-1$
with probability $\tilde{p}_{1,j,k}^p(f)$, respectively, and remains unpaired
with probability $\tilde{p}_{1,j}^u(f)$. 

Now, if base $j$ has been paired
to base $k$, recursively the sequence $r_1,\ldots,r_{k-1}$ is treated
in the same way
(\emph{case with force} $f$) 
and the sub sequence $r_{k+1},\ldots,r_{j-1}$ is treated
as described in the \emph{case without force}.

If base $j$ has not been paired, the sequence $r_1,\ldots,r_{j-1}$ is treated
in the same way (\emph{case with force} $f$).

\item \emph{Case without force} for sub sequence $r_i,\ldots,r_j$

Base $j$ is paired to one of the bases $k=i,\ldots,j-s-1$
with probability $p_{i,j,k}^p$, respectively, and remains unpaired
with probability $p_{i,j}^u$. 

 Now, if base $j$ has been paired
to base $k$, recursively the sequence $r_i,\ldots,r_{k-1}$ and
$r_{k+1},\ldots,r_{j-1}$ are treated in the same way 
(\emph{case without force}).

If base $j$ has not been paired, the sequence $r_1,\ldots,r_{j-1}$ is treated
in the same way (\emph{case without force}).

\end{itemize}

In this way, each time a structure is independently drawn according to the
Boltzmann distribution, i.e., the algorithm constitutes ideal sampling.

\subsection{Folding and Unfolding Algorithm}

The algorithm to determine the work
for a given sequence $\mathcal{R}$ works as follows:
First, a secondary structure is drawn in equilibrium
at some given initial value $f_0$ of the force and for RNA temperature $T$.
Then a Monte Carlo (MC) simulation
allowing to change the secondary structure with 
total of $n_{\rm MC}$ sweeps is performed while the force
parameter $f$ is increased or reduced depending on $\Delta f$. One
sweep consist of $L^2/2$ Monte Carlo steps.
During the MC simulation, $n_{\rm force}$ times the force is increased
by $\Delta f$.
For the individual MC steps,
each time two random residues $i$ and $j$ are selected.
If these are already paired
to each other, the pair is removed, i.e., the bond broken,
 with the usual Metropolis
 probability $p_{\rm Metr}=\min\{1,\exp(-\beta\Delta E)  \}$
 determined by the energy change $\Delta E = -e(r_i,r_j)$.
 Note that we use negative pair
 energies, thus  we have  always
 $p_{\rm Metr}=\exp(\beta e(r_i,r_j))$.
 In case of two non-bonded bases, they will be 
paired if they are complementary, and if they 
 have a distance larger than
$s$, and  if no
 pseudo-knots would be created. The configuration
 is not changed when just one of the
selected bases is already bounded, since a base can only connect to a
single other one. The random numbers which are used during the MC
simulation are generated before a call to the subroutine and stored
in a vector $\xi$. In this way, all the randomness is removed outside
this subroutine \cite{crooks2001}, for a reason we will present in the next
section. Note that all other parameters like $\mathcal{R}$, $T$ etc. remain
the same during a simulation, thus the work obtained during a unfolding or
refolding is   a deterministic function of $\xi$:

\begin{tabbing}
  \hspace{0.2cm} \= \hspace{0.2cm} \= \hspace{0.2cm} \=\\
  \textbf{algorithm} $W(\xi)$\\
  \textbf{begin}\\
  \> draw for $\mathcal{R}$ an equilibrium structure $\mathcal{S}$  at \\
  \>\>   initial force $f_0$ and RNA temperature $T$ \\
\> $f=f_0$ \\
\> $W=0$ \\
\> \textbf{for} $j=0,\cdots,n_{\rm force}$\\
\> \textbf{begin}\\
\> \> perform $L^2 n_{\rm MC}/(2n_{\rm force})$ MC-Steps:\\
\> \> \textbf{begin}\\
\> \> \> select two random residues $l,m\in \{1,\ldots,L\}$\\
\> \> \> \textbf{if} $(l,m) \in \mathcal{S}$, remove pair with prob. $p_{\rm Metr.}$\\
\> \> \> \textbf{else} if $(l,m)$ is allowed 
set $\mathcal{S}=\mathcal{S}\cup \left\lbrace (l,m)\right\rbrace$ \\
\> \> \textbf{end}\\
\> \> $f=f+\Delta f$\\
\> \> $W = W -n(\mathcal{S})\Delta f$\\
\> \textbf{end}\\
\> \textbf{return}($W$) \\
\textbf{end}	
\end{tabbing}

The vector $\xi=(\xi_1,\xi_2,\ldots,\xi_K)$ contains
$K=L-1 + 3L^2 n_{\rm MC}/2$
random numbers which are uniformly distributed in $[0,1]$.
These are all random numbers that are needed to
perform one full unfolding or refolding simulation.
Each random number has a specific fixed purpose.
The first
 $L-1$ entries are required to sample
an configuration from the partition function, where an individual
random number is utilized to determine if base $j\in[2,\dots,L]$ is
either connected to base $k\in[1,\dots,j-s-1]$ or unconnected. Not all
these
$L-1$ random numbers are necessarily used during a specific
sampling process, e.g.,
if for base $j$ the remaining sub sequence for a potential pairing partner
is too small. In this case, the corresponding random number is just ignored,
The subsequent MC steps need three random numbers each, two for selecting
a pair and potentially one more, if the Metropolis criterion is used.
If not, the third random number is also ignored, respectively.
This results
in a number of $3L^2n_{\rm MC}/2$ additional entries in $\xi$. 

Note that more efficient Monte Carlo algorithms for RNA secondary
structures exists \cite{flamm2000,dykeman2015}, which
are event-driven Gillespie algorithms. Also they
take as possible Monte Carlo moves only allowed
moves into account, i.e., either pairs are removed, or only allowed
pairs are proposed, avoiding non-complementary base pairs or pseudo knots.
This requires keeping track of the allowed moves, which also
generates quite some overhead in computation and it also involves 
the calculation of necessary corrections
factors due to the varying number of accessible neighboring secondary structure
configurations, in order
to guarantee detailed balance. Also, the Gillespie nature of these algorithms
make the use of random numbers dependent on the history of previous
events.
Nevertheless, for the present
application, the work process is embedded into another
higher-level Monte-Carlo simulation, see below. For 
a good performance of the higher-level MC
simulation this
 requires that for each entry of the vector a specific
 purpose is assigned, as presented above. if this requirement is met, small
 changes to $\xi$ yield typically small, i.e., not too ``chaotic''
 changes in the resulting work $W=W(\xi)$.
  This is the case with the present algorithm.

\subsection{Large-deviation approach}

Now we explain how the work simulations can be performed, such that
also the tails of the work distribution $P(W)$ with potentially
very small probabilities can be obtained. The method has been
introduced for the Ising model \cite{work_ising2014}, where more details
are given. Here we review only the general idea  and present the
specific details for our  study.

As mentioned in the previous section, for a given sequence $\mathcal{R}$,
temperature $T$ and the other parameters, which are all kept fixed for
a set of simulations, the outcome of the unfolding or refolding process
is solely determined by the random values contained in the vector $\xi$.
Thus, to perform a standard \emph{simple sampling} simulation, one could
each time draw a random vector $\xi$ with each entry being a pseudo random
number uniformly distributed in $[0,1]$. This results in one work value $W$
which sampled from the true distribution.
Thus, if one repeats the simple sampling many times, one can collect many
work values and calculate a histogram to approximate the full distribution.
Nevertheless, running the simple sampling $K$ times, will one only allow
to resolve probabilities larger or equal $1/K$ in the histogram.

In order to access to work distribution down to very small probabilities,
we did the following: We used a Markov chain Monte Carlo (MCMC)
simulation where
the states of the simulation are represented by samples $\xi^{(t)}$
of the random vectors used to drive the RNA unfolding or folding simulations.
Thus, each state of the Markov chain corresponds to exactly one instance
of a full
process consisting of starting with an initial state in equilibrium
and performing a, typically fast, non-equilibrium process during which
the force is changed, the system has a bit of time to relax between
two force changes, and a work
value $W=W(\xi^{(t)})$ is obtained in the end.
Therefore, the MCMC simulation takes place on a higher level than
the unfolding or refolding simulations.
Now, the main idea is
to include a bias in the MCMC simulation, which involves a Metropolis
acceptance depending on the change in the resulting work.

To be more precise, say we have the current state $\xi^{(t)}$ with work
$W^{(t)}=W(\xi^{(t)})$ in the MCMC
simulation. First, we generate a \emph{trial state} $\xi'$, which we obtain
by copying $\xi^{(t)}$ and then redrawing a number $n_{\xi}<K$ of
randomly selected entries from the $K$ entries of $\xi^{(t)}$.
Next, we perform a complete work process for $\xi'$, which results in the
measured work $W'=W(\xi')$. Now, the trial state is accepted, i.e.,
$\xi^{(t+1)}=\xi'$ with Metropolis probability
$\tilde p_{\rm Metr}=\min\{1,\exp(-\Delta W/\Theta)\}$, where
$\Delta W=W'-W^{(t)}$ is the change in work and  $\Theta$ is a temperature-like
control parameter. Otherwise, the trial state is rejected, i.e.,
$\xi^{(t+1)}=\xi^{(t)}$.
Note we aim at an empirical acceptance rate around 0.5
such that $n_{\xi}$ is typically small for small values of $\Theta$ and larger
for larger values of $\Theta$. Actual values are given below.

Since the setup of the MCMC simulation is like
  any standard MCMC approach for a system coupled to a heat bath,
  only that we have replaced the energy by the work and use $\Theta$
  for the temperature, it is obvious the our approach will sample
  the true work distribution but including a bias which is exactly
  the Boltzmann factor $\sim \exp(-W/\Theta)$.
  As usual, one has to discard the initial phase of the Markov chain,
  i.e., the equilibration phase,
  and to draw sample values only at suitable large time intervals.
  Thus, one can
  in principle perform simulations for a given value of $\Theta$,
  measure a histogram approximating the biased distribution
  $P_\Theta(W)\sim P(W) \exp(-W/\Theta)$, and obtain an estimate,
  up to the normalization constant, for
  the true distribution $P(W)$ by multiplication with $\exp(+W/\Theta)$.
  Note that, technically, to obtain the distribution over a large range
  of the support, one needs to perform simulations at several 
  suitably chosen values of the control temperature $\Theta$,
  obtain the normlization constants for all measured histograms
  and combine them
  into one single finally normalized
  histogram \cite{align2002}. Details, in particular
  for the case of the work distribution of on Ising model in an external
  field, can be found elsewhere \cite{work_ising2014}. This approach has
  already been applied to other non-equilibrium processes like the
  Kardar-Parisi-Zhang model \cite{kpz2018} or traffic flows
  \cite{nagel_schreckenberg2019}.

\subsection{Example secondary structures}

\begin{figure}[!t]
  \fbox{\includegraphics[width=\linewidth]{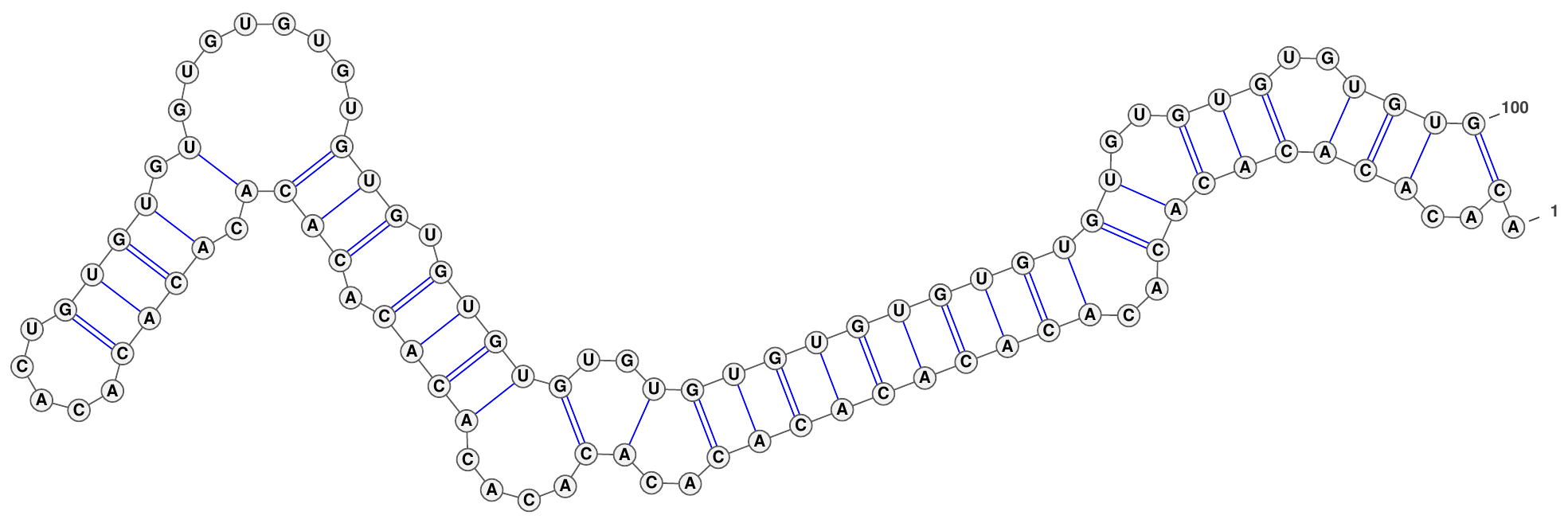}}\\
  \vspace{0.2cm}
  \fbox{\includegraphics[width=\linewidth]{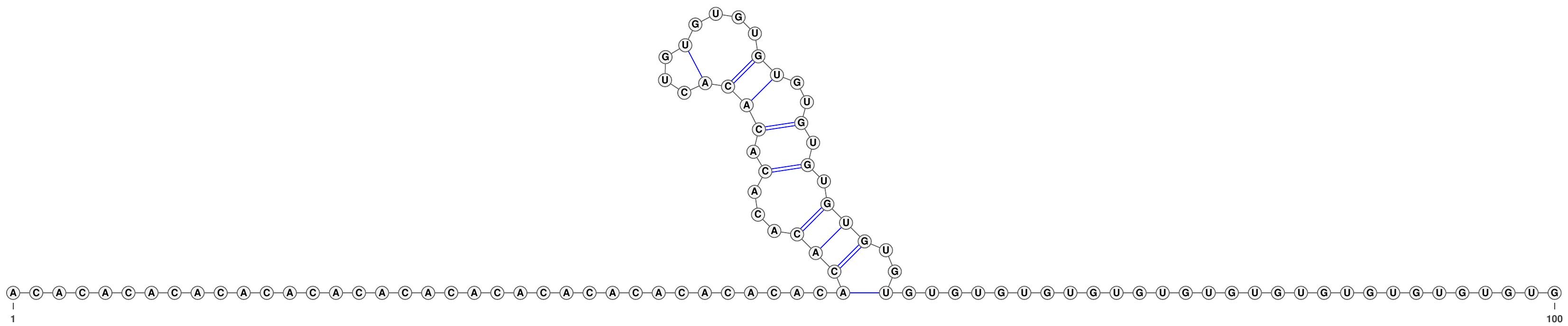}}\\
  \vspace{0.2cm}
  \fbox{\includegraphics[width=\linewidth]{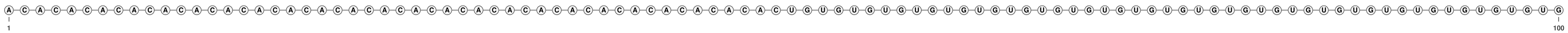}}
  \caption{\label{fig:secondary:structures} Exemplary equilibrium
    secondary structures at $T=1$ for different forces. Top: $f=0$.
    Middle: $f=0.805$. Bottom: $f=2$. Drawn with the VARNA package
    \cite{darty2009}.}
\end{figure}

In Fig.~\ref{fig:secondary:structures} equilibrium secondary structures
are shown.  It becomes apparent how the extension increases with 
the force parameter $f$.

\subsection{Simulation parameters}

For all unfolding and refolding processes, the force was increased
from $f_0=0$ to $f_{\max}=2$ and vice versa, with 400 steps each. Thus,
the change of the force was $\Delta f = \pm 0.005$.
Table \ref{tab:parameters} shows the other simulation parameters we have used.

\begin{table}[!h]
\begin{tabular}{rrcrrrrrr}
  $T$ & $n_{\rm MC}$ & $f$ &$n_{\Theta}$ & $\Theta_{\min}$ & $\Theta_{\max}$
  & $n_{\xi,\min}$ & $n_{\xi,\max}$ & $t_{\rm ld}/10^8$  \\ \hline
  0.3 & 8 & $0\to 2$ & 17 & 0.6 & 7 & 938 & $9\times 10^4$ & 5.44\\
  0.3 & 8 &  $2\to 0$ & 10 & 0.4 & 2 & 1587 & $6\times 10^4$ & 4.05\\
  0.3 & 16 & $0\to 2$ & 18 & 0.6 & 8 & 1407 & $12\times 10^4$ & 2.50\\
  0.3 & 16 & $2\to 0$ & 10 & 0.4 & 2 & 2500 & $9\times 10^4$ & 1.82\\
 1 & 8 & $0\to 2$ & 11 & 0.8 & 10 & 354 & $6\times 10^4$ & 6.95\\  
 1 & 8 & $2\to 0$ & 13 & 1 & 5 & 1350 & $6\times 10^4$ & 2.81\\  
 1 & 16 & $0\to 2$ & 11 & 0.8 & 10 & 938 & $75 \times 10^2$ & 4.41\\  
 1 & 16 & $2\to 0$ & 10 & 0.8 & 5 & 2344 & $12\times 10^4$ & 2.35 \\  
\end{tabular}
\caption{Simulation parameters for different temperatures $T$ and for
  different process speeds $n_{\rm MC}$ and unfolding ($f=0\to 2$)
  and refolding ($f=2\to 0$) processes. For the large-deviation MCMC
  simulation $n_{\Theta}$ different values of the temperature-like
  parameter $\Theta \in [\Theta_{\min},\Theta_{\max}]$ were
  considered. In each MCMC step a number
  $n_{\xi}\in [n_{\xi,\min},n_{\xi,\max}]$ of entries
  from the vectors $\xi$ of random numbers are changed. For the lowest value
  of $\Theta$ we have $n_{\xi} =n_{\xi,\min}$, for the largest
  $n_{\xi} =n_{\xi,\max}$, for the others in between. The total number
  of MCMC steps in the large-deviation simulation was always larger
  than the given values $t_{\rm ld}$, the actual values depending
  on the value of $\Theta$ and on
  the available computing time on the computing cluster, respectively.
  The longest running time occurred for the unfolding (forward)
  process  $T=1,n_{\rm MC}=8$ and took
  $t_{\rm ld}= 14.5\times 10^8$ steps.
  \label{tab:parameters}}
\end{table}

\subsection{Work distributions}

\begin{figure}[htb]
  \includegraphics[width=\linewidth]
                  {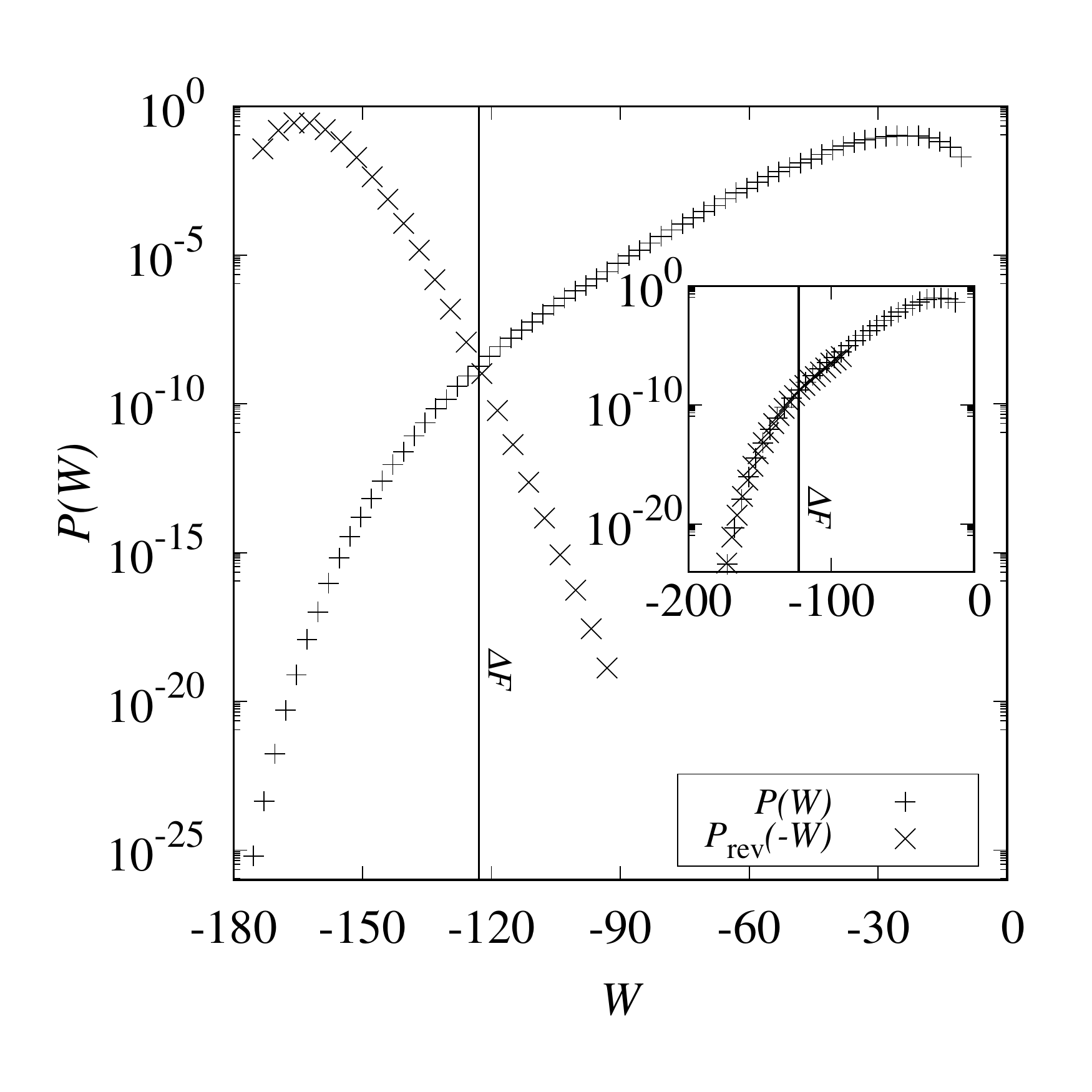}
     \caption{\label{fig:workdistribution:T1:16sweeps} Plain and mirrored
                    work distributions for
          $T=1$ and 16 sweeps of the forward and reverse process, respectively.
                    They intersect near $W=\Delta F$, which
          is the exact value and indicated by the vertical line.
          The inset shows the same plot but with the distribution for the
          reverse process (cross symbols) 
          rescaled as $P_{\rm rev}(W)\exp(-(\Delta F-W)/T)$,
          according to the equation of Crooks, yielding a good
                  agreement with $P(W)$.}
\end{figure}

In Fig.~\ref{fig:workdistribution:T1:16sweeps} the work distributions for $T=1$ 
of the slower process, at total of 16 MC sweeps per process, are shown.
The results look similar to the 8 sweeps case, but the
distributions are located a bit closer to each other here, such that the
intersections of $P(W)$ and $P_{\rm rev}(-W)$ occur at higher
probability.
In the inset of   Fig.~\ref{fig:workdistribution:T1:16sweeps} the corresponding
rescaled distribution for the reverse process is shown.
Also for 16 MC sweeps a good agreement with the distribution for the forward
process is visible, over more than 20 decades.

\begin{figure}[htb]
  \includegraphics[width=\linewidth]
                  {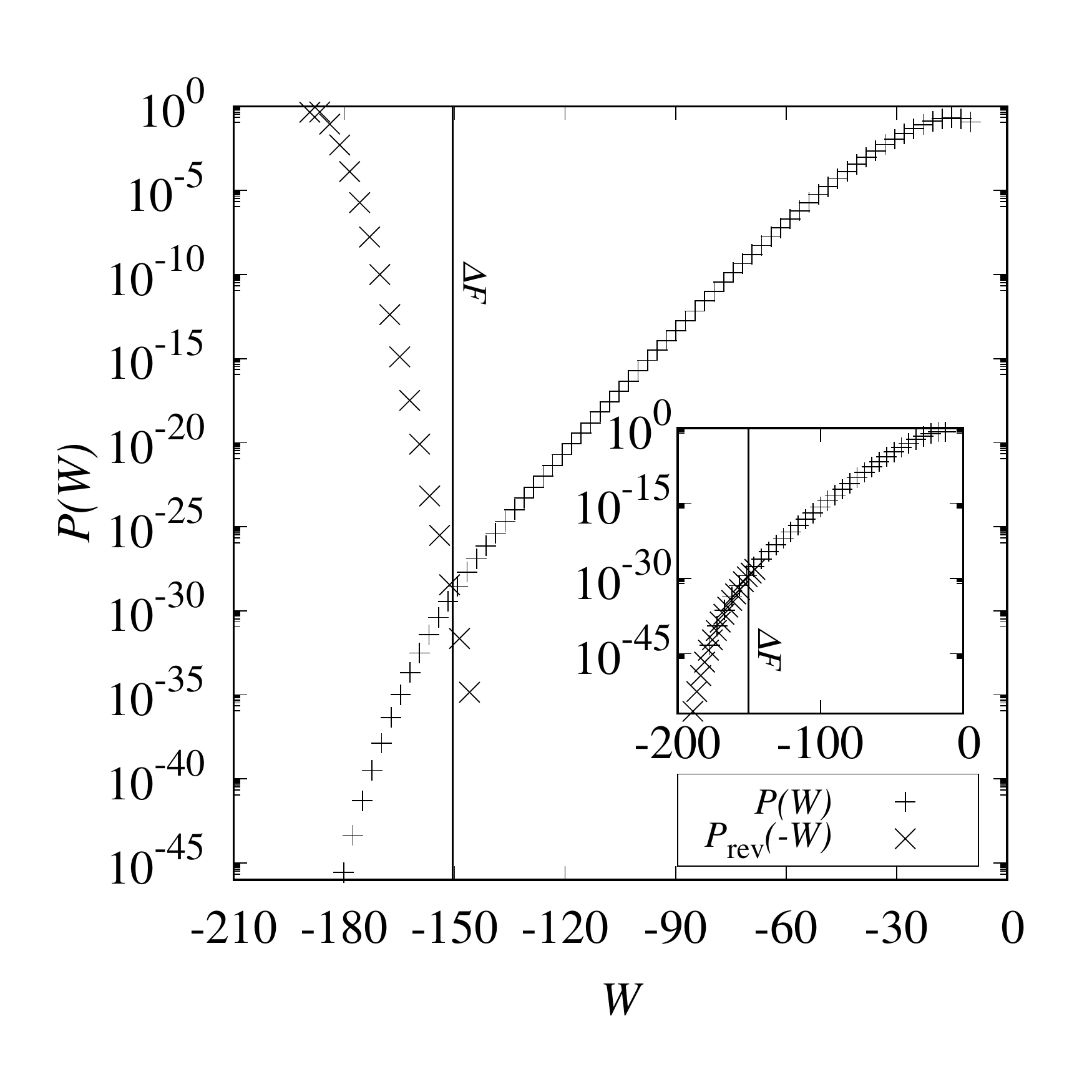}
     \caption{\label{fig:workdistribution:T03:8swees} Plain and mirrored
                    work distributions for
          $T=0.3$ and 8 sweeps of the forward and reverse process, respectively.
                    They intersect near $W=\Delta F$, which
          is the exact value and indicated by the vertical line.
          The inset shows the same plot but with the distribution for the
          reverse process (cross symbols) 
          rescaled as $P_{\rm rev}(W)\exp(-(\Delta F-W)/T)$,
          according to the equation of Crooks, yielding a good
                  agreement with $P(W)$.}
\end{figure}

\begin{figure}[htb]
  \includegraphics[width=\linewidth]
                  {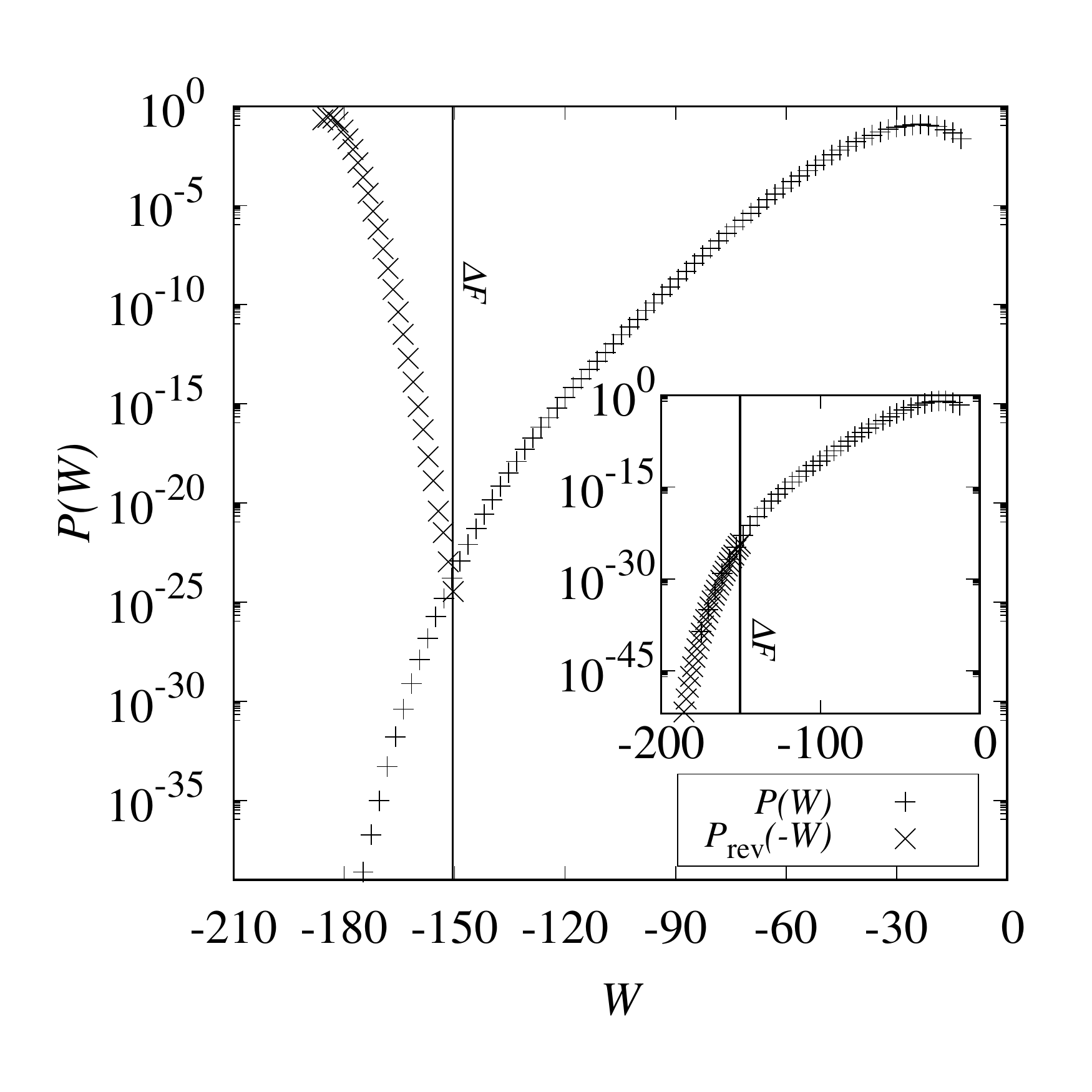}
     \caption{\label{fig:workdistribution:T03:16swees} Plain and mirrored
                    work distributions for
                    $T=0.3$ and 16 sweeps of the forward and reverse process,
                    respectively.
                    They intersect near $W=\Delta F$, which
          is the exact value and indicated by the vertical line.
          The inset shows the same plot but with the distribution for the
          reverse process (cross symbols) 
          rescaled as $P_{\rm rev}(W)\exp(-(\Delta F-W)/T)$,
          according to the equation of Crooks, yielding a good
                  agreement with $P(W)$.}
\end{figure}

In Figs.~\ref{fig:workdistribution:T03:8swees} and
\ref{fig:workdistribution:T03:16swees} the corresponding results
for the lower temperature $T=0.3$ are shown. Again, Crooks theorem
is confirmed with high precision, where here the distribution was even obtained
down to probabilities as small as $10^{-46}$.

\subsection{Jarzynski Integrand}

\begin{figure}[!ht]
  \includegraphics[width=\linewidth]{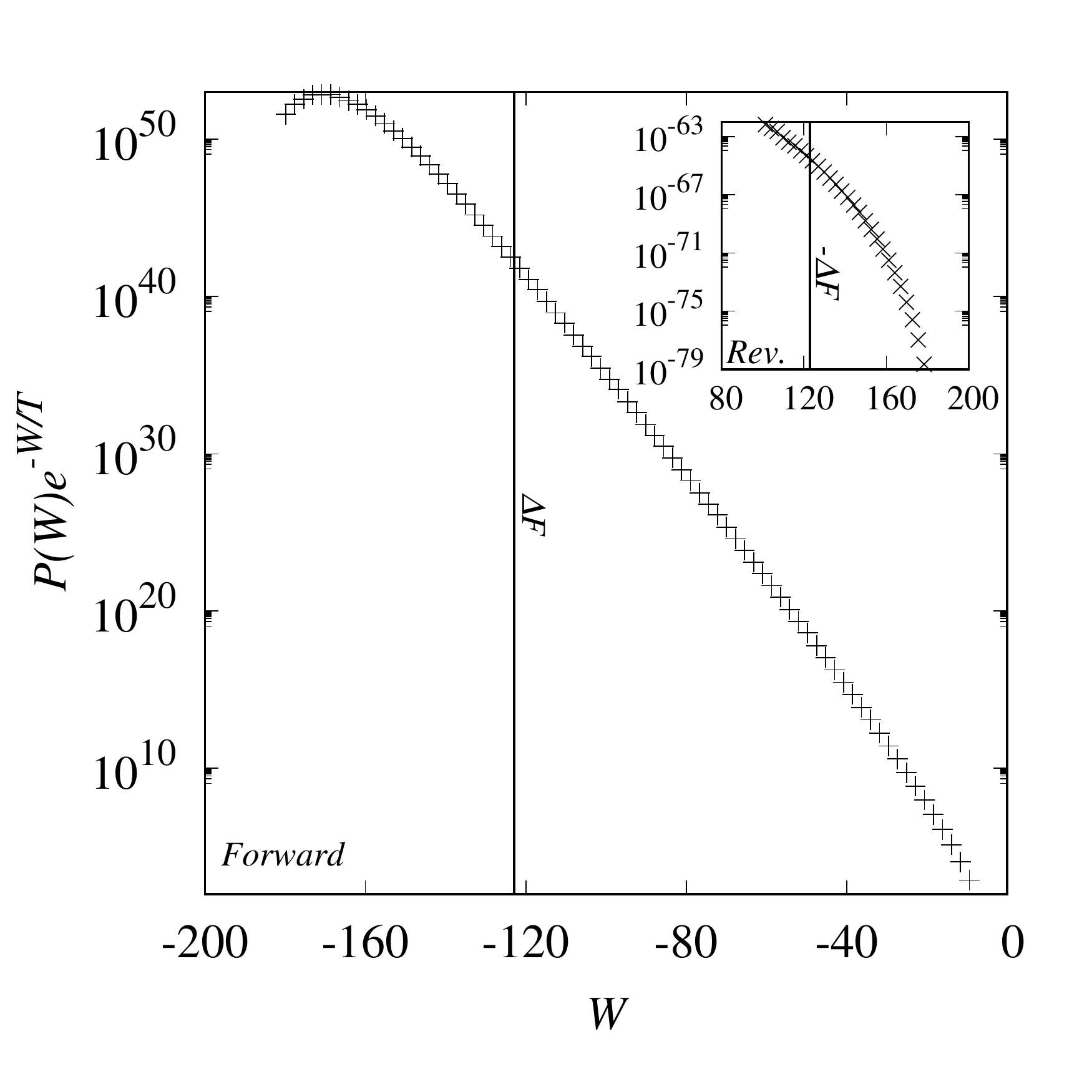}
  \caption{\label{fig:jarzynski:integrand} Jarzynski integrand of
    the forward process for 8 sweeps at $T=1$.
    Inset: Same for the reverse process, in which the maximum is not
    reached. Error bars are smaller than symbol sizes. }
\end{figure}

The integrand of $\langle e^{-W/T}\rangle=\int dW P(W) e^{-W/T}$
is shown in Fig.\ \ref{fig:jarzynski:integrand}, for $T=1$,
$n_{\rm MC}=8$ and forward and reverse work processes, respectively.
The point where the integrand peaks is exponentially relevant and can
be used  to approximate the integral. This, together with its
probability, 
determines according to Jarzynski's equation the free energy difference.

\subsection{Overlap}

For two secondary structures $\mathcal{S}$ and $\mathcal{S'}$ and the
equivalent notations $\{s(i)\}$ and $\{s'(i)\}$ for the pairing
partners  of the residues (0 if not paired),
we define the \emph{overlap}

\begin{equation}
  \sigma(\mathcal{S},\mathcal{S}') = \frac 1 L \sum_{i=1}^L
  \delta_{s(i),s'(i)}\,
  \end{equation}
  where the Kronecker delta is given by $\delta_{k,l}=1$ if $k=l$ and
  $\delta_{k,l}=0$ else. Thus, the overlap equals one when 
  $\mathcal{S},\mathcal{S}'$ denote the same secondary structure,
  and zero when they are completely different.

\subsection{Force-extension curves}

In addition to the overlap profiles $\sigma(f)$ we have presented
in the main paper, we also used
force extension curves (FECs) $n(f)$ to compare processes
for equilibrium and non-equilibrium situations. Note that the extension
$n(\mathcal{S})$ of a structure can be very much influenced by single
base pairs. Thus two processes, which look very similar on the
level of secondary structures, can be very different with respect
to force-extension curves.

\begin{figure}[htb]
  \includegraphics[width=0.49\linewidth]{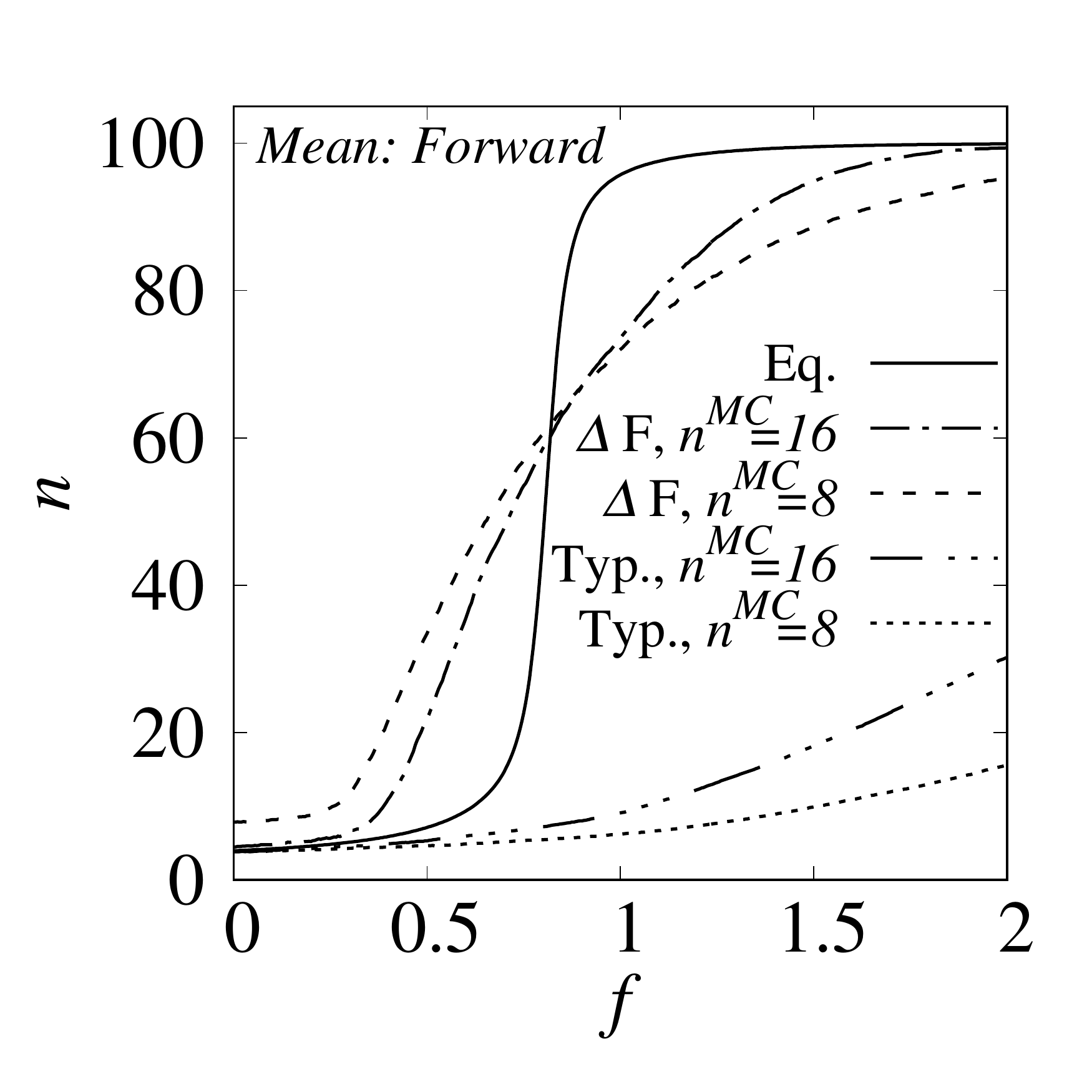}
  \includegraphics[width=0.49\linewidth]{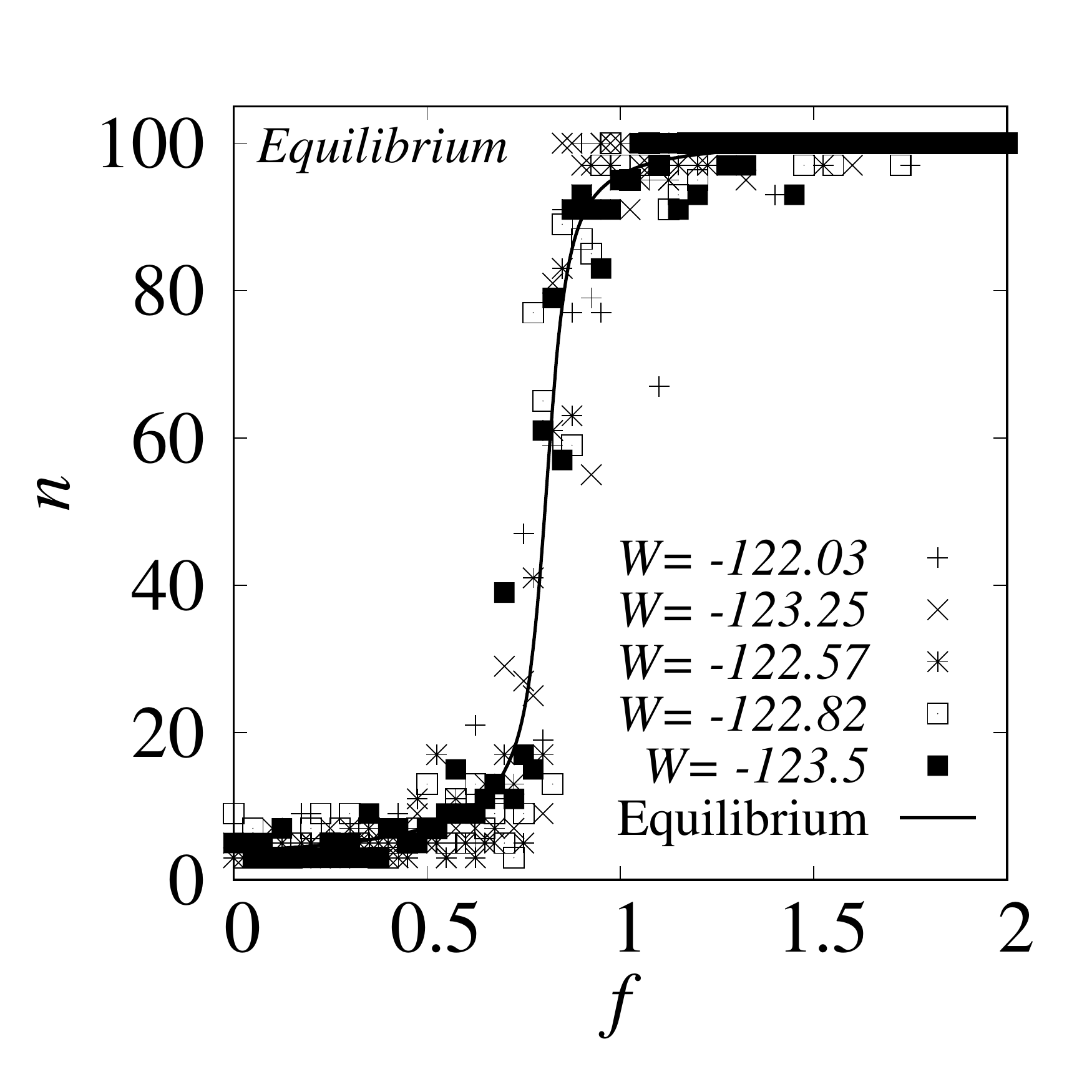} 
  \includegraphics[width=0.49\linewidth]{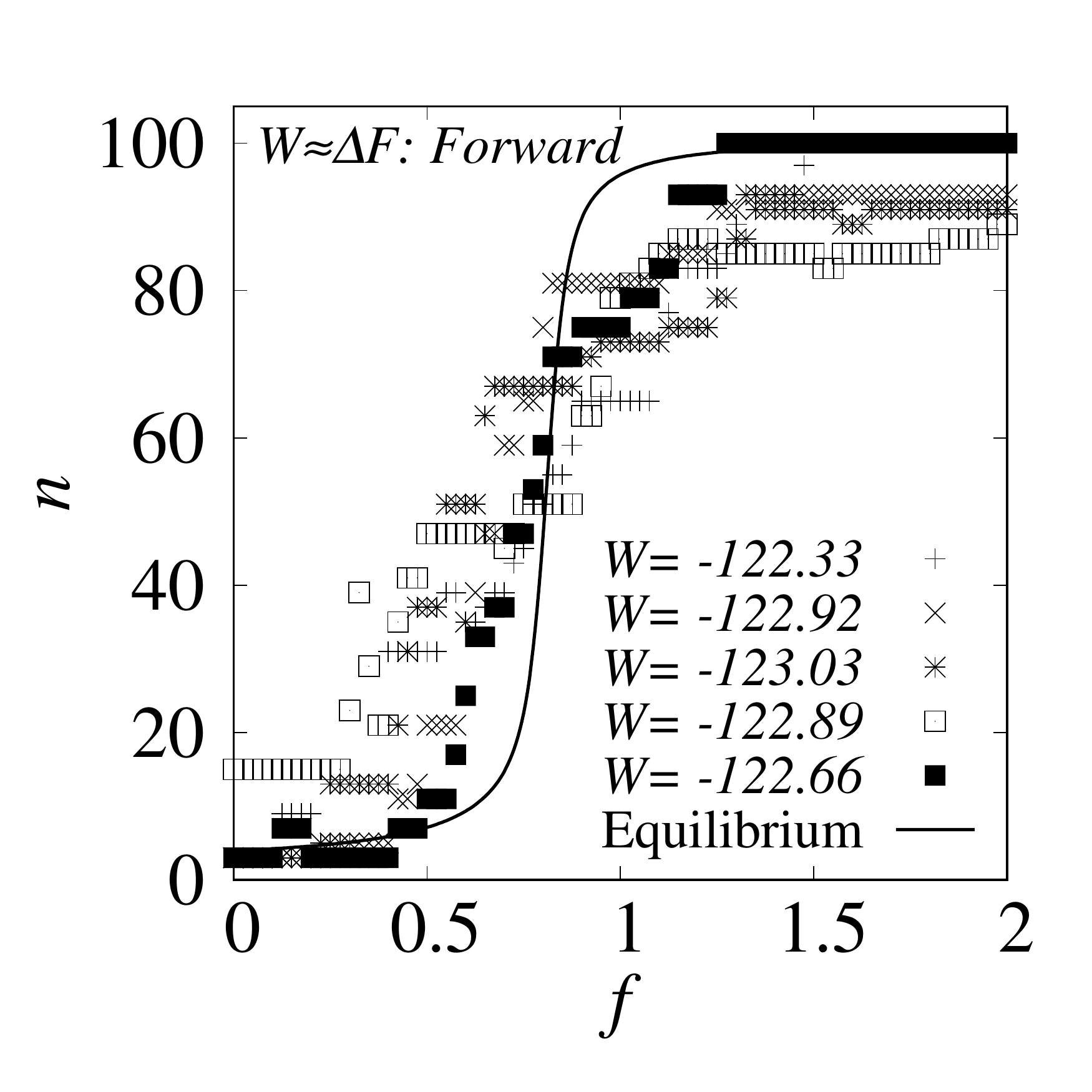}
  \includegraphics[width=0.49\linewidth]{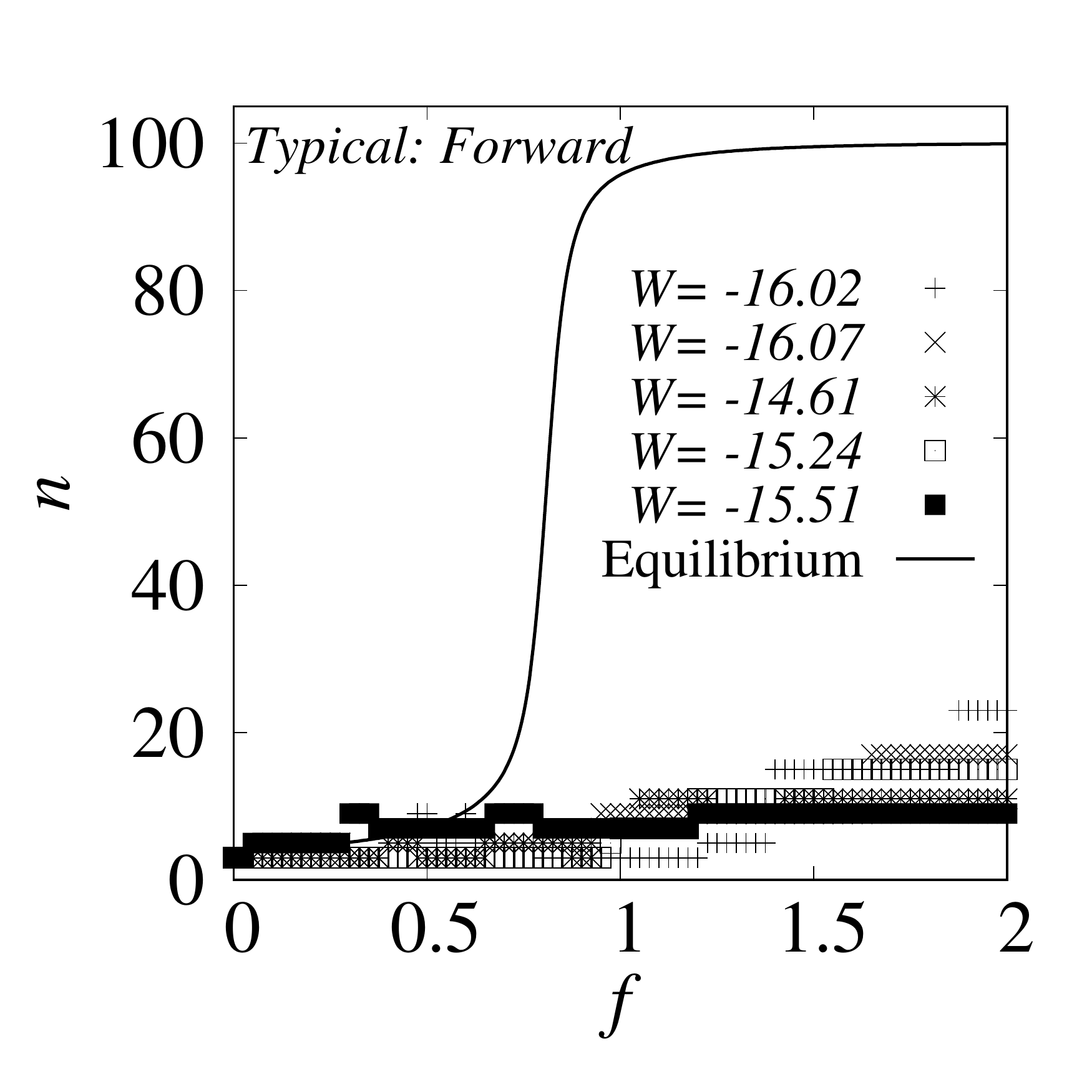}
  \caption{\label{fig:force:extension:curves}
    Top left: Mean FECs, in equilibrium,
    for typical forward processes and for work values near
    $\Delta F$, for two different number of $n_{\rm MC}$ of sweeps at $T=1$.
    Top right: Samples of such single FECs in equilibrium.
    Bottom left: samples of non-equilibrium FECs with $n_{\rm MC}=8$
    for $W$ near $\Delta F$.
    Bottom right: samples of typical  non-equilibrium FECs, i.e.,
    where $W\gg \Delta F$, with $n_{\rm MC}=8$.
   The solid line represents always the mean equilibrium  FEC. }
\end{figure}

Samples for equilibrium and non-equilibrium FECs for forward processes,
along with corresponding averages, are shown in
Fig.~\ref{fig:force:extension:curves}.
For the equilibrium case, a sigmoidal form can be observed, with
some fluctuations, with a strong change near the critical force
value where the folding-unfolding transition takes place \cite{mueller2002}.
For the non-equilibrium case, the typical FECs, i.e., with typical
work values $W$ far from $\Delta F$,
agree only for small values of $f$, i.e., in the
initial phase of the process. On the other hand, the rare processes with
$W$ close to $\Delta F$, where five different examples
are shown here, are much more similar to the equilibrium FECs. Here,
differences appear mainly
near the critical folding-unfolding force.

\begin{figure}[!b]
  \includegraphics[width=0.49\linewidth]{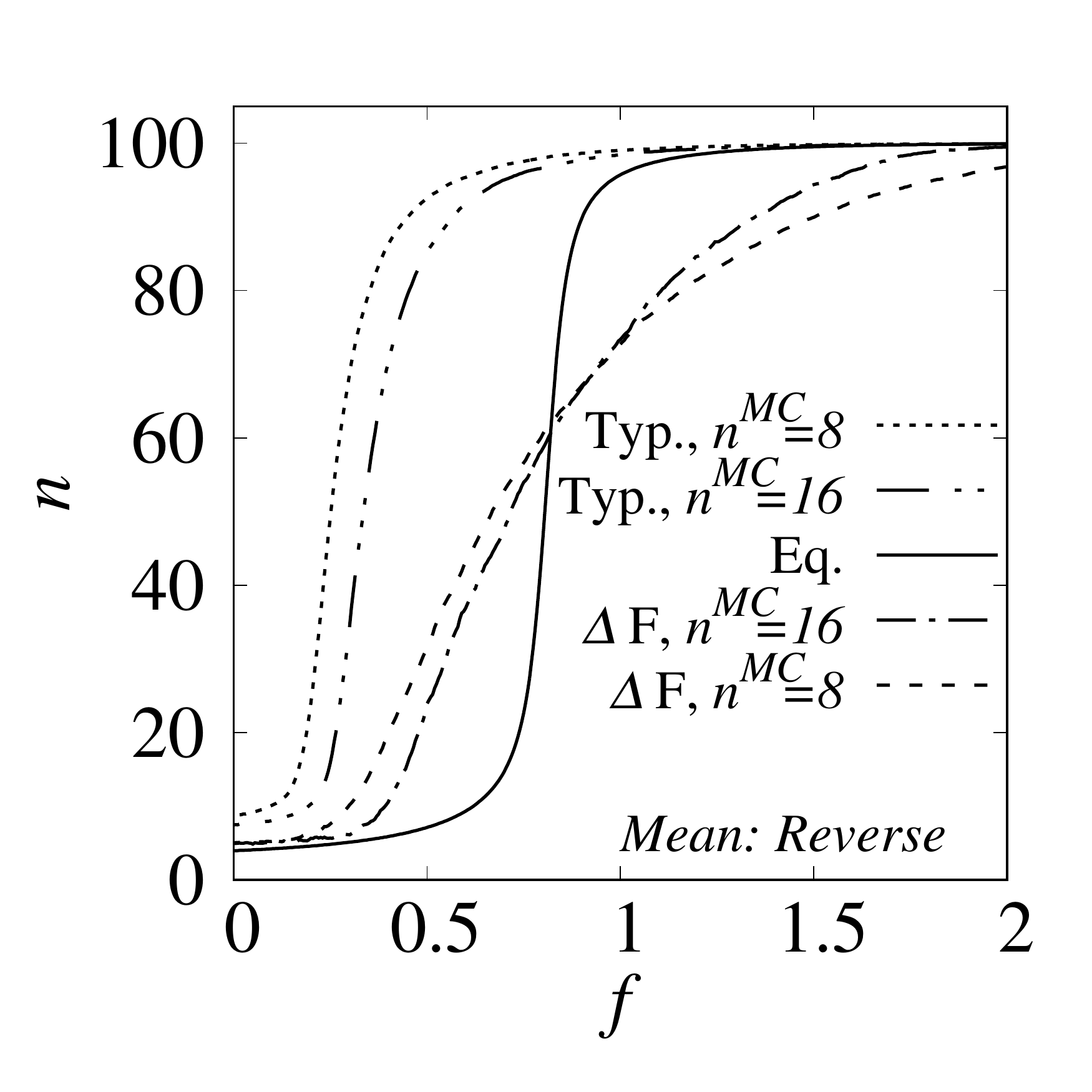}
  \includegraphics[width=0.49\linewidth]{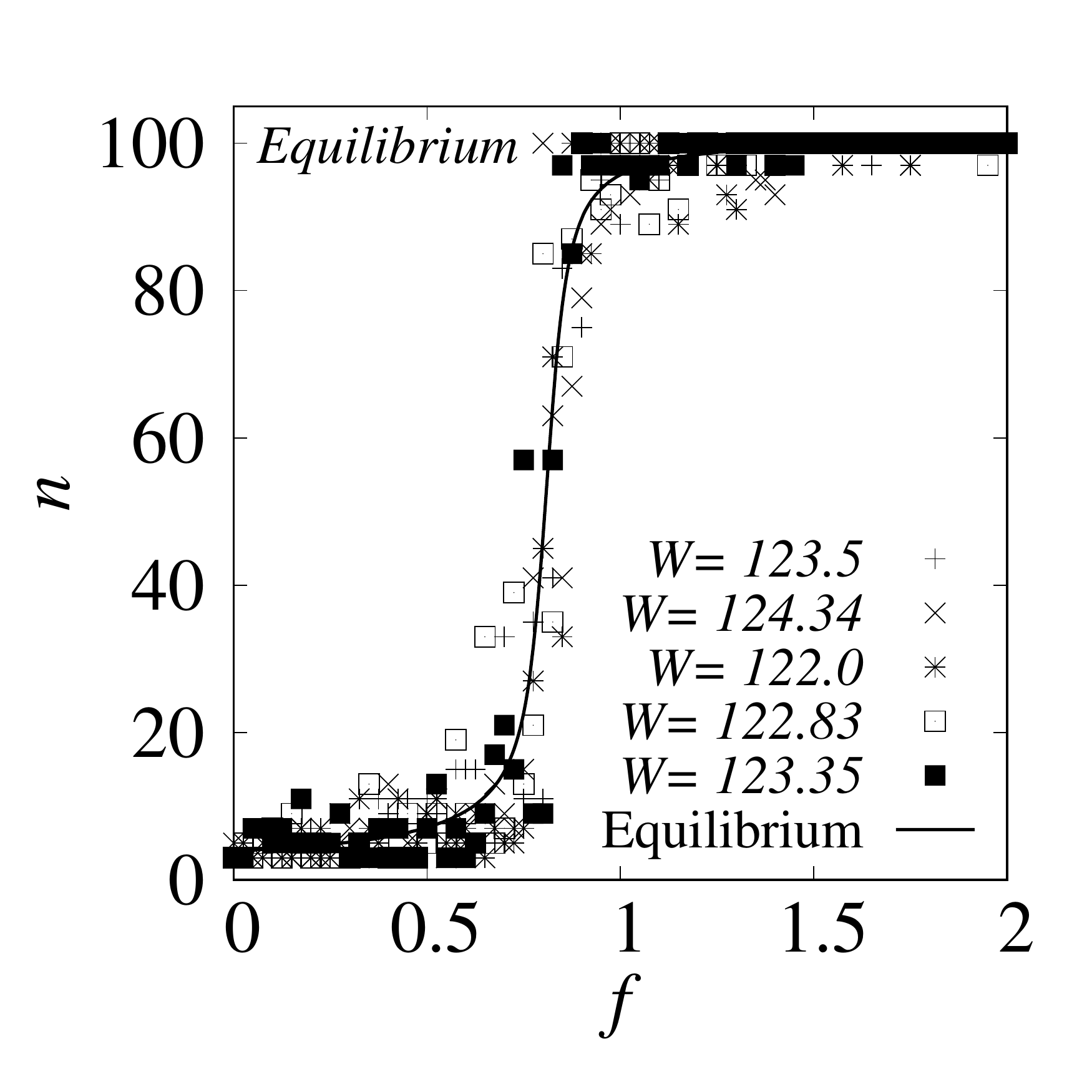}
  \includegraphics[width=0.49\linewidth]{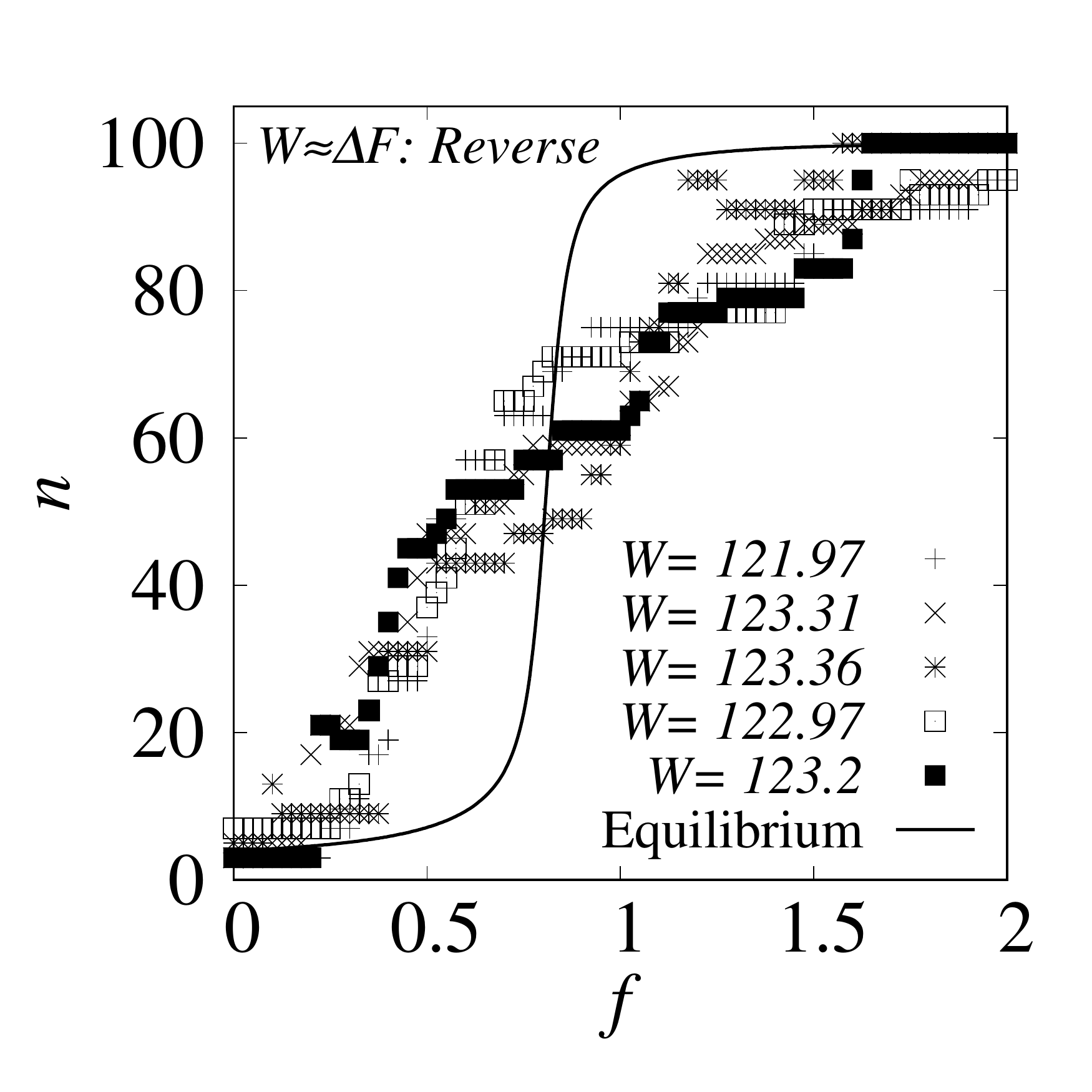}
  \includegraphics[width=0.49\linewidth]{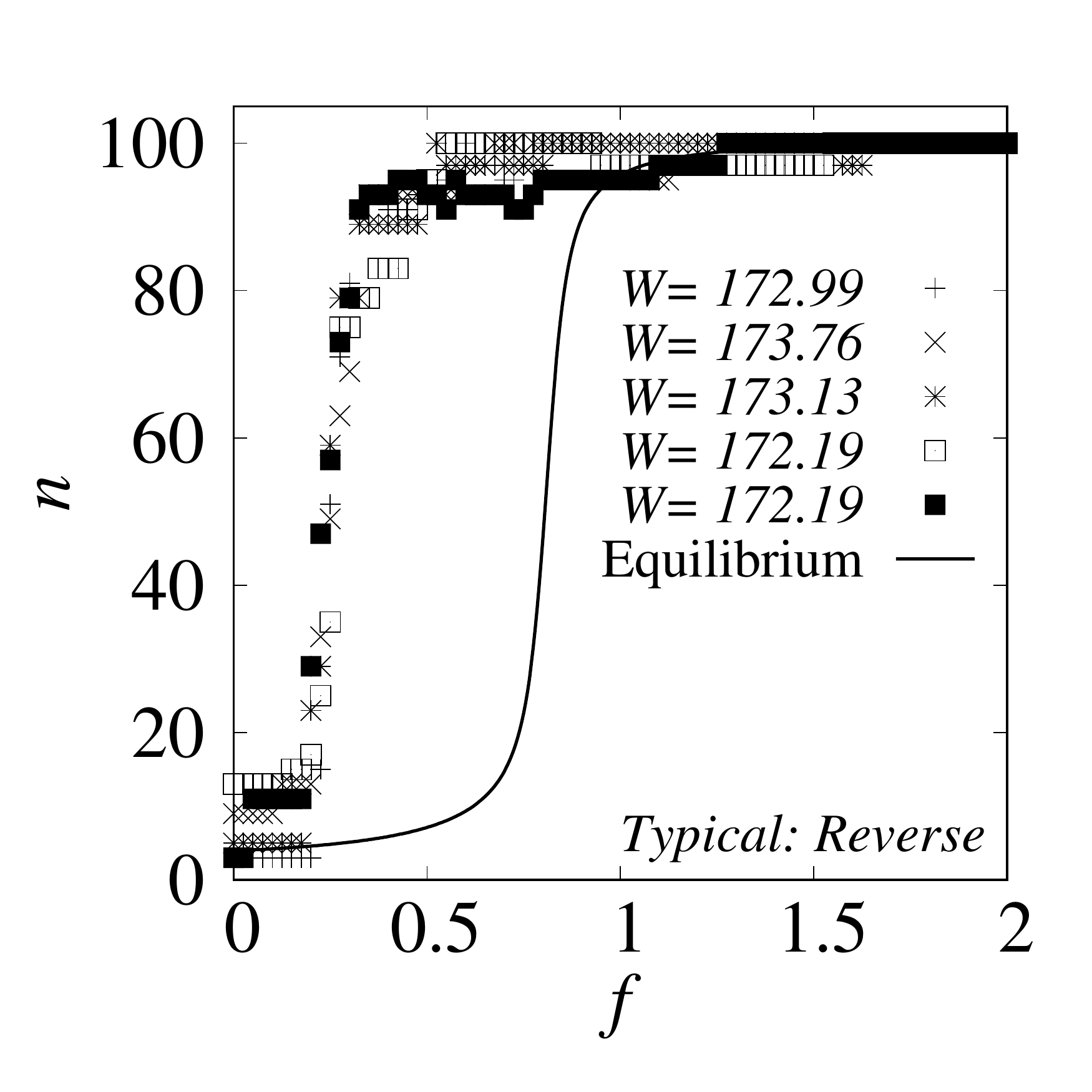}
  \caption{\label{fig:force:extension:curves:reverse}
    Top left: Mean FECs, in equilibrium,
    for typical reverse processes and for work values near
    $\Delta F$, for two different number of $n_{\rm MC}$ of sweeps at $T=1$.
    Top right: Samples of such single FECs in equilibrium.
    Bottom left: samples of non-equilibrium reverse
    FECs with $n_{\rm MC}=8$ for $W$ near $\Delta F$.
    Bottom right: samples of typical  non-equilibrium reverse FECs, i.e.,
    where $W\gg \Delta F$, with $n_{\rm MC}=8$.
   The solid line represents always the mean equilibrium  FEC.
 }
\end{figure}

Samples for equilibrium and non-equilibrium FECs for backward processes,
along with corresponding averages, are shown in
Fig.~\ref{fig:force:extension:curves:reverse}. The results correspond to the
forward case, but the processes with \emph{typical} values of $W$
agree well with the
average equilibrium FEC only for large values of $f$
but not for small values of $f$. But this means they also agree  in the
initial phase of the process, before the critical folding-unfolding
force value is reached. The FECs for work values $W\approx \Delta F$  are
also for reverse processes much more similar to the equilibrium case
than typical reverse processes.

\begin{figure}[htb]
	\includegraphics[width=\linewidth]{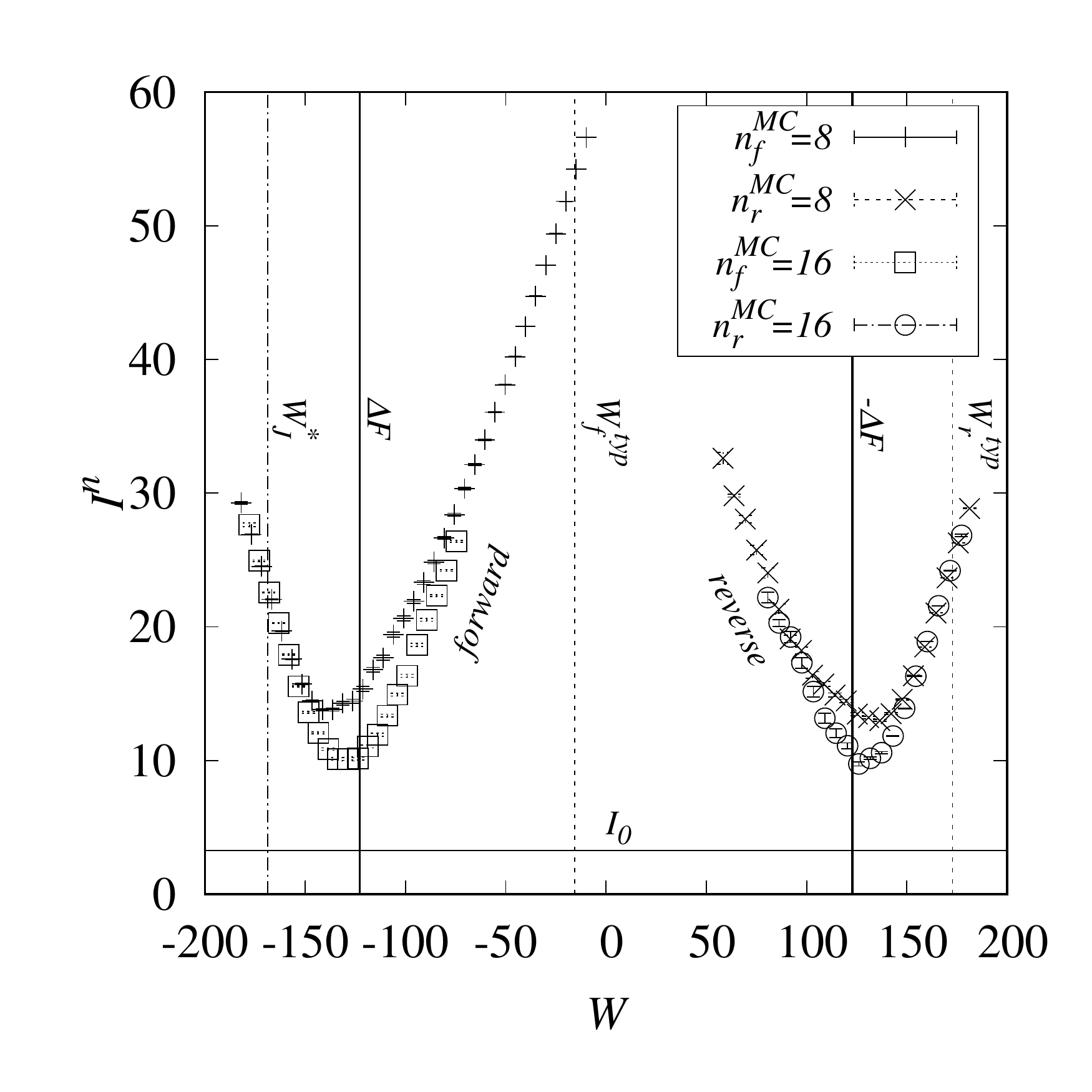}
	\caption{\label{fig:integrated:free:length:difference}
          Integrated extension difference $I^{n}$ between equilibrium
          and non-equilibrium at $T=1$. For 8 sweeps the entire work
          range is plotted, where for 16 sweeps only a range around
          the minimum is shown, for better visibility.
          $I_0$, represented by a horizontal
          line, is the averaged value of $I^{n}$ when comparing
          always two equilibrium FECs. The left
          curves represent the forward, the right ones the reverse
          process. Vertical lines indicate work values at (from left
          to right) the maximum $W_J^*$ of the Jarzynski integrand ,
          the free energy difference $\Delta F$, the maximum $W^{typ}_f$ of the
          forward process work distribution , the negative
          free energy difference $-\Delta F$ and the maximum point
          $W^{typ}_r$ of
          the reverse process work distribution. }
\end{figure}

These results are confirmed by averaging the absolute value of the
differences between one FEC $n(f)$ and the mean equilibrium FEC $\overline{n}_{Eq}(f)$ over all available values of the force $f$,
i.e., calculating $I^n=[\frac 1 {n_{\rm f}} \sum_f |n(f)-\overline{n}_{Eq}(f)|]$  where the average $[\ldots ]$ is over different
realisations of $n(f)$. Even when considering equilibrium
FECs for $n(f)$, respectively,
there is some variation reflected by a non-zero average value $I_0$.
When using non-equilibrium FECs, with a specified binned value of $W$,
one sees stronger differences,
as visible in Fig.~\ref{fig:integrated:free:length:difference}.
Similar to $I^{\sigma}$, the closest
agreements between non-equilibrium and equilibrium are seen near
$W\approx \Delta F$. In contrast to $I^{\sigma}$ the level
of the equilibrium fluctuations is not reached for the measurable quantity FEC.

\bibliography{lit}

\end{document}